\documentclass{aa}

\usepackage{graphicx, comment}
\usepackage{txfonts}
\usepackage[colorlinks=true]{hyperref}
\hypersetup{
     colorlinks   = true,
     citecolor    = blue
}

\usepackage{float}
\usepackage{placeins}

\newcommand{\C}{\object{3C\,84}}

\newcommand{\qfour}{$\beta^{Q_4}_\textrm{app} = 0.116\pm0.008$c}
\newcommand{\qfourmasyr}{$\mu^{Q_\textrm{4}} = 0.094\pm0.007\,$mas/yr}

\newcommand{\qsix}{$\beta^{Q_6}_\textrm{app} = 0.115\pm0.012$c}
\newcommand{\qsixmasyr}{$\mu^{Q_\textrm{6}} = 0.093\pm0.010\,$mas/yr}

\newcommand{\wten}{$\beta^{W_{10}}_\textrm{app} = 0.013\pm0.013$c} 
\newcommand{\wtenmasyr}{$\mu^{W_\textrm{10}} = 0.011\pm0.011\,$mas/yr} 

\newcommand{\wfourteen}{$\beta^{W_{14}}_\textrm{app} = 0.078\pm0.042$c} 
\newcommand{\wfourteenmasyr}{$\mu^{W_\textrm{14}} = 0.063\pm0.034\,$mas/yr}

\newcommand{\qcomb}{$\beta^{Q_\textrm{comb}}_\textrm{app} = 0.116\pm0.001$c}
\newcommand{\qcombmasyr}{$\mu^{Q_\textrm{comb}} = 0.094\pm0.001\,$mas/yr}

\newcommand{\gammaonefive}{$0.36\pm0.38$}
\newcommand{\gammafourthree}{$0.27\pm0.13$}
\newcommand{\gammaeightsix}{$0.90\pm0.06$}
\newcommand{\Gammaonefive}{$\sim0.1$}
\newcommand{\Gammafourthree}{$\sim0.3$}
\newcommand{\Gammaeightsix}{$\sim0.4$}
\newcommand{\medff}{0.091\pm0.027}
\newcommand{\medfe}{0.082\pm0.024}
\newcommand{\medffnoise}{0.099\pm0.030}
\newcommand{\medfenoise}{0.085\pm0.021}

\newcommand{\qaver}{$\beta^{Q_\textrm{avg}}_\textrm{app} = 0.087\pm0.032$c}
\newcommand{\qavermasyr}{$\mu^{Q_\textrm{avg}} = 0.071\pm0.026\,$mas/yr} 

\newcommand{\waver}{$\beta^{W_\textrm{avg}}_\textrm{app} = 0.16\pm0.06$c}
\newcommand{\wavermasyr}{$\mu^{W_\textrm{avg}} = 0.13\pm0.05\,$mas/yr}

\newcommand{\wavertwt}{$\beta^{W^{2010}_\textrm{avg}}_\textrm{app} = 0.18\pm0.04$c}
\newcommand{\wavertwtmasyr}{$\mu^{W^{2010}_\textrm{avg}} = 0.15\pm0.03\,$mas/yr} 

\newcommand{\comaver}{$\beta^\textrm{avg}_\textrm{app} = 0.055-0.22$c} 
\newcommand{\comavermasyr}{$\mu^\textrm{avg} = 0.04-0.18\,$mas/yr}

\begin{document}

    \title{Jet kinematics in the transversely stratified jet of \C}
   \subtitle{A two-decade overview}

   \author{G.~F. Paraschos\inst{1}, T.~P. Krichbaum\inst{1}, J.-Y. Kim\inst{12, 2, 1}, J.~A. Hodgson\inst{3, 2}, J. Oh\inst{3}, E. Ros\inst{1}, J.~A. Zensus\inst{1}, A.~P. Marscher\inst{4}, S.~G. Jorstad\inst{4,5}, M.~A. Gurwell\inst{6}, A. L\"ahteenm\"aki\inst{7,8}, M. Tornikoski\inst{7}, S. Kiehlmann\inst{9,10}, A.~C.~S. Readhead\inst{11}
          }
   \authorrunning{G. F. Paraschos et al.}
   \institute{$^{1}$Max-Planck-Institut f\"ur Radioastronomie, Auf dem H\"ugel 69, Bonn, D-53121 Bonn, Germany\\ 
              $^{}$\ \email{gfparaschos@mpifr-bonn.mpg.de}\\
              $^{2}$Korea Astronomy and Space science Institute, 776 Daedeokdae-ro, Yuseong-gu, Daejeon, 30455, Korea\\
              $^{3}$Department of Physics and Astronomy, Sejong University, 209 Neungdong-ro, Gwangjin-gu, Seoul 05006, Korea\\
              $^{4}$Institute for Astrophysical Research, Boston University, Boston, MA 02215, USA\\
              $^{5}$Astronomical Institute, Saint Petersburg State University, Universitetsky Prospekt, 28, Petrodvorets, 198504 St. Petersburg, Russia\\
              $^{6}$Center for Astrophysics | Harvard \& Smithsonian, 60 Garden Street, Cambridge, MA 02138, USA\\
              $^{7}$Aalto University Mets\"ahovi Radio Observatory, Mets\"ahovintie 114, 02540 Kylm\"al\"a, Finland\\
              $^{8}$Aalto University Department of Electronics and Nanoengineering, P.O. BOX 15500, FI-00076 AALTO, Finland.\\
              $^{9}$Institute of Astrophysics, Foundation for Research and Technology-Hellas, GR-71110 Heraklion, Greece\\
              $^{10}$Department of Physics, Univ. of Crete, GR-70013 Heraklion, Greece\\
              $^{11}$Owens Valley Radio Observatory, California Institute of Technology, Pasadena, CA 91125, USA\\
              $^{12}$Department of Astronomy and Atmospheric Sciences, Kyungpook National University, Daegu 702-701, Republic of Korea
             }

   \date{Received -; accepted -}

  \abstract{
  \C\ (\object{NGC\,1275}) is one of the brightest radio sources in the mm radio-bands, which led to a plethora of VLBI observations at numerous frequencies over the years.
  They reveal a two-sided jet structure, with an expanding but not well-collimated parsec-scale jet, pointing southward.
  High resolution mm-VLBI observations allow the study and imaging of the jet base on the sub-parsec scale. 
  This could facilitate the investigation of the nature of the jet origin, also in view of the previously detected two-railed jet structure and east-west oriented core region seen with {\it RadioAstron} at 22\,GHz.
  We produce VLBI images of this core and inner jet region, observed during the past twenty years at 15, 43, and 86\,GHz. 
  We determine the kinematics of the inner jet and ejected features at 43 and 86\,GHz and compare their ejection times with radio and $\gamma$-ray variability. 
  For the moving jet features, we find an average velocity of \comaver\ (\comavermasyr). 
  From the time-averaged VLBI images at the three frequencies, we measure the transverse jet width along the bulk flow. 
  On the $\leq 1.5$\,parsec scale,  we find a clear trend of the jet width being frequency dependent, with the jet being narrower at higher frequencies.
  This stratification is discussed in the context of a spine-sheath scenario, and is compared to other possible interpretations.
  From quasi-simultaneous observations at 43 and 86\,GHz we obtain spectral index maps, revealing a time-variable orientation of the spectral index gradient, due to  structural variability of the inner jet.
}

   \keywords{
            Galaxies: jets -- Galaxies: active -- Galaxies: individual: 3C\,84 (NGC\,1275) -- Techniques: interferometric -- Techniques: high angular resolution
               }

   \maketitle

\section{Introduction}
Studying the kinematics of AGN jets at the sub-parsec scale has only recently been made possible by the advancement of technology utilised in centimetre- and millimetre very-long-baseline interferometry (VLBI). 
Still, only a few sources provide the necessary criteria required to resolve the complex, physical phenomena at play. 
The best suited candidate AGN should be nearby, harbour a super massive black hole, which powers a jet, preferably transversely resolved, and which is oriented with respect to the observer at a moderately large viewing angle to avoid strong Doppler beaming effects. 
The existence of measurements over a long timeline at high cadence, and at different frequencies, is also advantageous.

\C, the radio source associated with the central galaxy \object{NGC\,1275} in the Perseus cluster \citep{Ho97}, fulfils all these criteria.
The prominent type 2 radio-galaxy is indeed very nearby ($D_L = 78.9$\,Mpc, $z=0.0176$, \citealt{Strauss92})\footnote{We assume $\Lambda$ cold dark matter cosmology with $H_0 = 67.8$\,km/s/Mpc, $\Omega_\Lambda = 0.692$, and $\Omega_M=0.308$ \citep{Planck16}.} and harbours a super massive black hole (M$_{\rm BH} \sim 9 \times 10^8$\,M$_\odot$ , \citealt{Scharwaechter13}). 
A multitude of epochs at numerous frequencies is available in the literature, as the source has been studied in detail over more than half a century \citep{Baade54, Walker94, Dhawan98, Walker00, Suzuki12, Nagai14, Dutson14, Giovannini18, Kim19, Paraschos21}.
Two jets emanate from the central engine \citep{Vermeulen94, Walker94, Fujita17}; one moves southward and is very prominent, while the northern one is much fainter and is obscured by free-free absorption on milli-arcsecond (mas) scales \citep{Wajima20}. 
The subject of the exact value of the viewing angle is still in contention. VLBI studies at 43\,GHz indicate a viewing angle of $\theta_\textrm{LOS}\approx65^\circ$ \citep{Fujita17} based on the jet to counter-jet ratio; on the other hand, spectral energy distribution (SED) fitting of $\gamma$-rays place the viewing angle on the lower side, at $\theta_\textrm{LOS}\approx25^\circ$ \citep{Abdo09} or even $\theta_\textrm{LOS}\approx11^\circ$ \citep{Lister09b}.
We note that the different viewing angles are measured in different regions and therefore  suggest a spatial bending of the jet.
Nevertheless, the viewing angle is large enough to allow for excellent tracking of the different radio emitting features \citep{Oh22}.
The jet kinematics derived from observing these features over multiple VLBI maps yield rather slow speeds, accelerating from $\beta_\textrm{app}\leq 0.1c$ on sub-mas scales up to $\sim1.4c$ on mas-scales \citep{Krichbaum92, Dhawan98, Suzuki12, Punsly21, Hodgson21, Weaver22}.

In this work we present a comprehensive tracking and cross-identification of travelling features visible in the central 0.5\,mas jet region of \C, both at 43 and 86\,GHz, and spanning a time range of more than two decades. 
Tracing the individual travelling features over such a time range provides valuable insight into the physical mechanisms in action at a relatively small distance 
to the super massive black hole. 
The component position registration at different frequencies bears the potential
to study the stratification of the inner jet.

Furthermore, we look at the expanding and collimating behaviour of the jet, by adding images (stacking process) which are observed close in time (quasi-simultaneous) at 15, 43, and 86\,GHz. 
From this we obtain an estimate of the initial expansion profile within the first milli-arcsecond radius from the VLBI core.
We also discuss in this context the pressure power-law index derived from the expansion profile and the implications for the medium surrounding the inner jet.

This paper is organised as follows: In Sec. \ref{sec:Data} we summarise the observations and data reduction; in Sec. \ref{sec:Results} we present our analysis and results. 
In Sec. \ref{sec:Disc} we discuss our results and in Sec. \ref{sec:Conclusions} put forward our conclusions.

\section{Observations and data reduction}\label{sec:Data}
\subsection{Data description and total intensity calibration} \label{sec:Data1}

We have analysed all available 86\,GHz VLBI observations of \C, which make a total of twenty-four full-track VLBI experiments, ranging from April of 1999 to October of 2020, which to our knowledge are all available epochs of \C\ at the particular frequency observed with the Coordinated Global Millimeter VLBI Array (CMVA) and its successor, the Global Millimeter VLBI Array (GMVA).
The observations were made at a cadence of roughly one per year.
The superior resolution that 3\,mm VLBI offers, provides a unique opportunity to study the parsec-scale jet of \C\ in great detail.
The post-correlation analysis was done in a similar manner as described in \cite{Kim19}, using the standard procedures in \textsc{AIPS} (\citealt{Greisen90}; cf. \citealt{MartiVidal12}).

Furthermore, we obtained eighteen VLBI images of \C\ from the \emph{VLBA-BU-BLAZAR} monitoring program, at 43\,GHz. 
Further details of the structure of the program, as well as a description of the data reduction and results can be found in \cite{Jorstad05} and \cite{Jorstad17}.
The earliest epochs available are from 2010; we chose one or two epochs per year, up to the most recent one (May 2021), to match the cadence of the 86\,GHz epochs.

The same selection procedure was followed for the 15\,GHz images. In this case we used the \emph{Astrogeo VLBI FITS} image database\footnote{\url{http://astrogeo.org/vlbi_images/}}.
We point out here that the images comprising this database are of heterogeneous quality, given the nature of observations. 
Since the earliest available epoch is dictated by the lack of 43\,GHz images before 2010, the first 15\,GHz epoch analysed is from 2010, reaching again up to 2020. 
An overview of the behaviour of \C\ at 15\,GHz is also presented in \cite{Britzen19}, where the authors analysed epochs between 1999 and 2017.  

A complementary element to our analysis are the radio variability, total flux light curves of \C\ at different frequencies.
For this investigation, we obtained radio light curves 4.8, 8.0, and 14.8\,GHz, ranging from 1980 to 2020, from the University of Michigan Radio Observatory (UMRAO), at 15\,GHz from the Owen's Valley Radio Observatory \citep[OVRO; see also][for a description of the observations and the data reduction]{Richards11}, at 37\,GHz from the Mets\"ahovi Radio Observatory (MRO) and at 230, and 345\,GHz from the Submillimeter Array (SMA).
Additionally, we used the publicly available\footnote{\url{https://fermi.gsfc.nasa.gov/ssc/data/access/lat/msl_lc/source/NGC_1275}} $\gamma$-ray light curve of \object{NGC\,1275} (\C), at MeV-GeV energies, which were carried out in survey mode (\citealt{Atwood09}; see also \citealt{Kocevski21}).
Detailed studies of the $\gamma$-ray emission in \C\ are 
available in the literature, for example, in \cite{Abdo09, Aleksic12, Nagai12, Nagai16, Hodgson18, Linhoff21, Hodgson21}. In this paper we focus on possible correlations of the $\gamma$-ray variability with the VLBI component ejection.
Finally, we used three historical measurements of the total intensity flux of \C, from the Mauna Kea Observatory (MKO) at 1.1\,mm \citep{Hildebrand77}, the National Radio Astronomy Observatory (NRAO) at 1.3\,mm \citep{Landau83}, and at the NASA Infrared Telescope Facility (IRTF) at 1.0\,mm \citep{Roellig86}.

\subsection{Imaging} \label{sec:Data2}

\begin{figure*}
   \centering
   \includegraphics[scale=0.35]{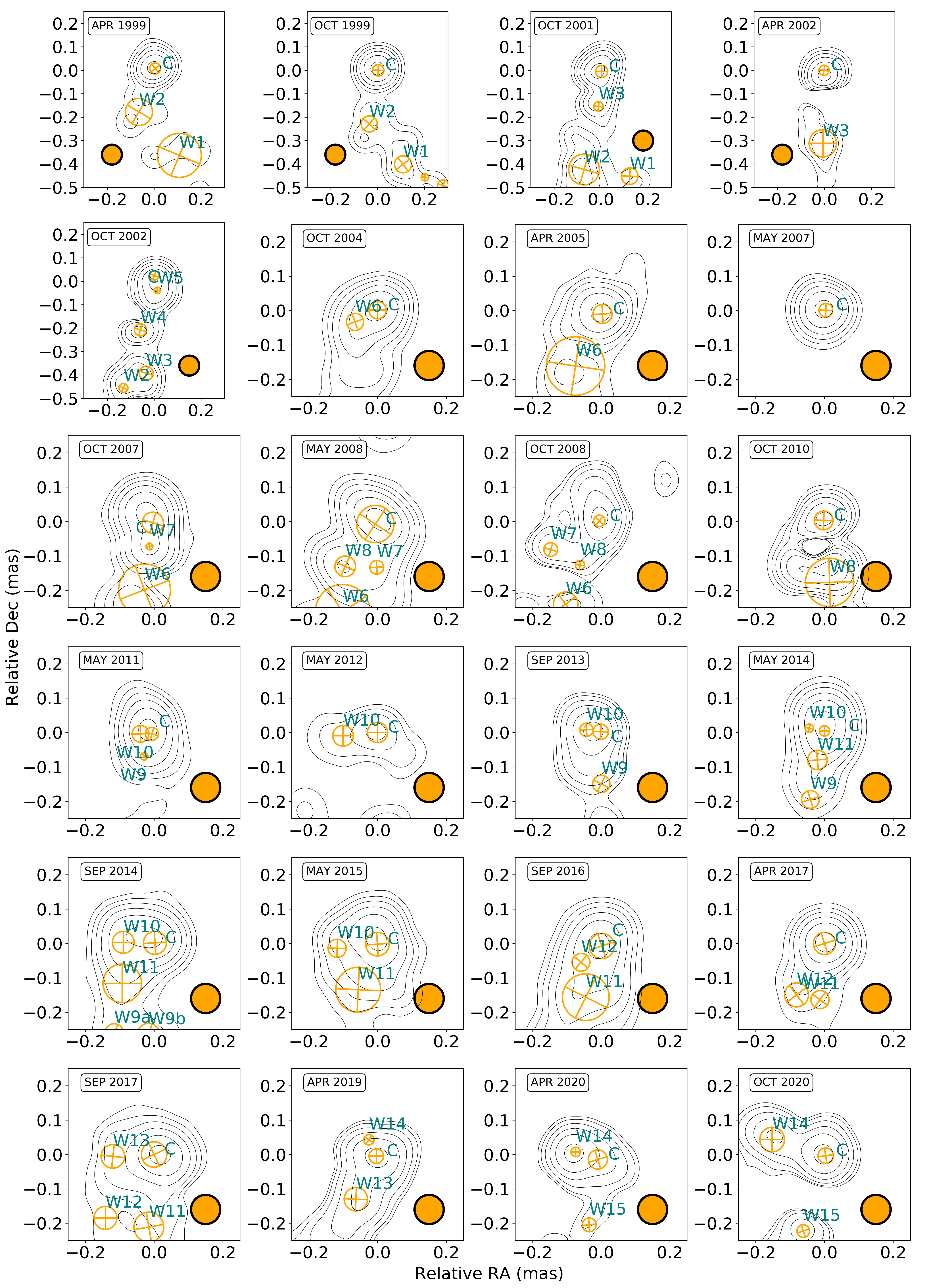}
   \caption{Circular Gaussian components super-imposed on 86\,GHz \textsc{clean} images. The legend of each panel indicates the epoch at which the observations were obtained. The beam of each epoch is displayed in the panels as a filled orange circle. 
   The contour levels correspond to 3\%, 5\%, 10\%, 20\%, 40\%, and 80\% of the intensity peak for all epochs, except May 2012, where 10\%, 20\%, 40\%, and 80\% of the intensity peak were used, because the data quality was limited.
   The epochs are convolved with the same circular beam, with a radius of 0.088\,mas.
   This radius size was obtained by geometrically averaging the beam sizes of all displayed epochs.
   A summary of the beam parameters of each individual epoch and peak fluxes is provided in Table~\ref{table:Beam86}.
   Each feature is labelled based on its date of emergence, with the lower number indicating an earlier ejection time. 
   As a centre of alignment we always used the north-westernmost feature inside the core region, which we labelled as ``C''.
   We note that feature ``W2'' was detectable in the April 2002 epoch and feature ``W6'' was not detectable in the May 2007 epoch, due to SNR limitations.
   A summary of the characteristics of each feature is provided in Table~\ref{table:Par86}.
   } 
    \label{fig:86mod}
    \end{figure*}

After having obtained the frequency averaged 86\,GHz data through fringe fitting and calibration, we analysed them with the \textsc{difmap} package \citep{Shepherd94}. 
We followed again the procedure described in \cite{Kim19} to obtain \textsc{clean} component (CC) images. 
The second step of our imaging procedure was to fit circular Gaussian components to the visibilities, which provides a  fit to the data with best possible $\chi^2$, and until the CC image was satisfactorily reproduced.
In our analysis, we omit Gaussian components, which represent the more extended emission beyond $r\geq 0.5$\,mas.
Their physical interpretation would be more ambiguous, owing to SNR limitations and their low brightness temperature.
Figure \ref{fig:86mod} displays the contour maps of the \textsc{CLEAN} images with circular Gaussian model fit components super-imposed as orange circles.
Details of the contour levels and rms noise magnitude are given in the caption.
The individual beam sizes and pixel scales are summarised in Table~\ref{table:Beam86}.

VLBI imaging and Gaussian component model fitting can be subjective. 
Some alternative methods have been utilised recently, such as wavelet-based analysis \citep[e.g.][]{Mertens15} and CC cluster analysis techniques \citep[e.g.][]{Punsly21}.
Inspired by the latter and in order to increase the robustness of our results, we implemented as a third step, an additional CC cluster analysis, which yields very similar results, within the error budgets, for the feature identification and their velocities as computed from the Gaussian component modelfitting (see Sec. \ref{sec:XID}). In view of uv-coverage limitations and residual calibration effects, we regard the latter method as most reliable, based on the smaller number of free fit parameters.

\begin{figure*}
   \centering
   \includegraphics[scale=0.35]{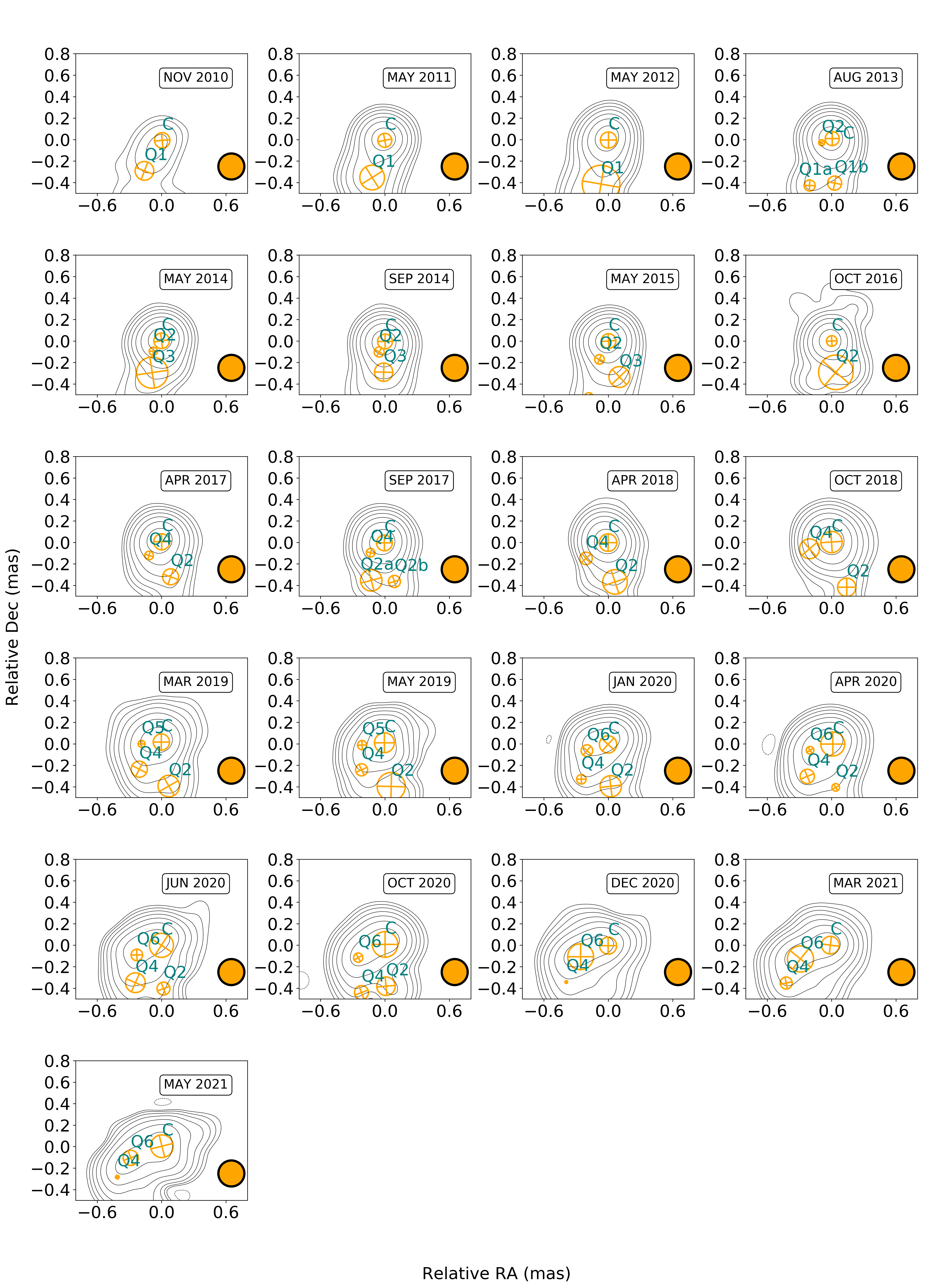}
   \caption{Circular Gaussian components super-imposed on 43\,GHz \textsc{clean} images. The legend of each panel indicates the epoch at which the observations were obtained. The beam of each epoch is displayed in the bottom right side as a filled orange circle. 
   The contour levels correspond to -1\%, 1\%, 2\%, 4\%, 8\%, 16\%, 32\%, and 64\% of the intensity peak, except in November of 2010, where, due to lower data quality, the contour levels were adjusted to -10\%, 10\%, 20\%, 40\%, and 80\% of the intensity peak.
   The epochs are convolved with the same circular beam, with a radius of 0.24 mas. 
   This size was obtained by geometrically averaging the beam sizes of all displayed epochs.
   A summary of the beam parameters and peak fluxes is provided in Table~\ref{table:Beam43}.
   Each features is labelled based on its date of emergence, with the lower number indicating an earlier ejection time. 
   As a centre of alignment we always used the north-westernmost feature inside the core region, which we labelled as ``C''.
   We note that feature ``Q1'' seems to split into two (labelled ``Q1a'' and ``Q1b'') in August 2013; similarly ``Q2'' seems to split into two (labelled ``Q2a'' and ``Q2b'') in September 2017. 
   These splits might be due to improved sensitivity in these epochs and/or variations of the intrinsic brightness and size of the jet feature.
   A summary of the characteristics of each feature is provided in Table~\ref{table:Par43}.
   } 
    \label{fig:43mod}
    \end{figure*}

For the 15 and 43\,GHz maps, we downloaded the respective data files (already fringe fit and calibrated) from the websites mentioned in Sec. \ref{sec:Data1}.
We then followed the same steps as with the 86\,GHz data sets.
Figure \ref{fig:43mod} showcases the circular Gaussian components, super-imposed on the CC images, and the individual beam sizes and pixel scales are summarised in Table~\ref{table:Beam43} for the 43\,GHz epochs.
The circular Gaussian component model fitting for the 15\,GHz data was exclusively done to align the images to each other, as in some epochs the C3 region \citep{Nagai14} was brighter than the nucleus.
Cross-identifying features at this frequency with the higher frequency maps is virtually impossible, due to the large beam size at 15\,GHz.
The beam sizes and pixel sizes are summarised in Table~\ref{table:Beam15}.

\section{Analysis and results}\label{sec:Results}
\subsection{Component cross-identification} \label{sec:XID}

Figures \ref{fig:86mod} and \ref{fig:43mod} showcase the travelling features labelled with identifiers (IDs), through the years.
Our procedure for identifying a travelling feature through the epochs was as follows: starting from the 43\,GHz maps, we established the trajectories of the main travelling features by requiring that the fit to their velocity produced the minimal dispersion. 
Continuity arguments also dictate that features cannot move upstream, barring some uncertainty in their position.
A formal description of our positional error estimation is presented in Appendix \ref{App:ParTab}.
Nevertheless, we note that some features have been reported to move back and forth (e.g. \citealt{Hodgson21}), although further downstream, in the C3 region.
Such a flip is interpreted as new features emerging.
The model parameters are presented in Table~\ref{table:Par43}.

We establish the trajectories of the core features at 86\,GHz.
Of course, one should expect to identify more features at 86GHz than at lower frequencies, since the generally smaller beam size resolves more fine structure near the core.
A comprehensive table of the model parameters is shown in Table~\ref{table:Par86}.
Five of the six 43\,GHz features could be robustly cross-identified with 86\,GHz features, based on matching position, time of appearance, and velocity.
We labelled these features as in Table~\ref{table:Vel4386}.
We note here, however, that our analysis implicitly assumes that the features are first visible at 86\,GHz near the core and only later become visible at 43 GHz, when they have travelled further downstream.
It can be argued that a scenario, in which the ejection near the core is not picked up at 86\,GHz due to the sparse cadence, and is instead only later detected downstream at 43\,GHz, is also possible.
Finally we point out, that for the cross-identification, we only focused on the positional information provided by the visibility phases, and not the absolute flux information, since the flux density at 86\,GHz can be subject to large uncertainties. 
The relative fluxes, however, are less uncertain.

In order to render results more robust, we used the CC cluster method for estimating feature velocities as follows: first we super-resolved each image, with a beam size at least half of the nominal value presented in Tables \ref{table:Beam86}, \ref{table:Beam43}, and \ref{table:Beam15}, to visually recognise each bright feature location, corresponding to a CC cluster.
Then we computed the coordinates of the CC cluster centroid by calculating the average of the CC coordinates in the image plane, inside a radius the size of the corresponding modelled Gaussian component.
We used the \textit{jackknife technique} \citep{Efron81} to estimate the mean position.
For the above described procedure, we only considered CC components with a flux higher than 5\% of the maximum flux CC inside the radius, defined as described above.
This approach produced a second set of feature coordinates, allowing us to estimate their velocities more confidently (see Sec.~\ref{sec:Kin}), by averaging these values and the ones obtained by fitting the standard Gaussian component coordinates.

\subsection{Kinematics} \label{sec:Kin}

\begin{figure*}[ht]
   \centering
   \includegraphics[scale=0.3]{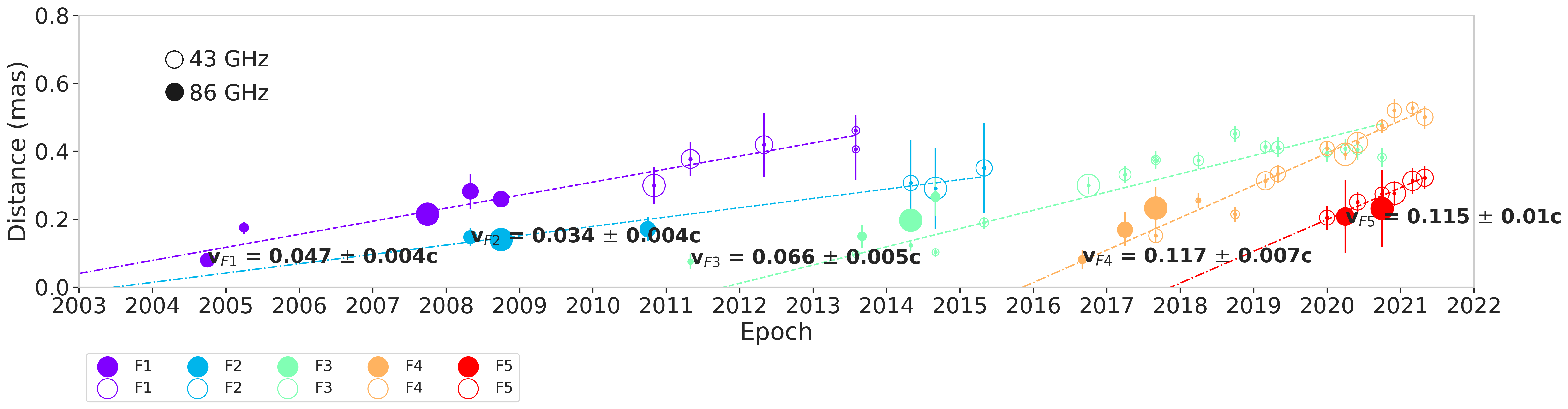}\\
      \caption{Plot of the motion of the cross-identified, colour-coded features. The empty and filled symbols correspond to the 43 and 86\,GHz features respectively. The velocity fit with the uncertainty is displayed above the first occurrence of each feature.
      The size of each symbol is normalised to the flux of the central feature (``C'') of that epoch.
      The velocities are also listed in Table~\ref{table:Vel4386}.}
         \label{fig:Vel_Comb}
   \end{figure*}

In Fig.~\ref{fig:Vel_Comb} we show the cross-identified features. 
Each travelling feature is colour-coded and the open and filled symbols correspond to the 43 and 86\,GHz travelling features respectively.
The dashed lines describe the velocity and (back-extrapolated) temporal point of ejection.
The velocities and their uncertainties of the cross-identified features ``F1'', ``F2'', ``F3'', ``F4'', and ``F5'' are listed next to the first time each feature is identified.
They are calculated from a weighted averaging of the velocities obtained from the Gaussian component positions and from the CC cluster centroids.

Having established the feature correspondence through the years, we performed a weighted linear regression analysis to each trajectory.
A formal description of the fitting procedure is presented in Appendix \ref{App:Comp_vel_mod}.
Figure \ref{fig:Vel_Comb} displays the apparent motions of the travelling features.
We find that the average velocity of the 86\,GHz features amounts to \wavermasyr\ (\waver), whereas for 43\,GHz, the features move on average at \qavermasyr\ (\qaver).
In order to compare the two, we also compute the 86\,GHz velocity of features ejected after 2010.
This calculation yields an average velocity of \wavertwtmasyr\ (\wavertwt) at 86\,GHz\footnote{These values are obtained by computing the weighted mean of the data points, with the weights being the inverted squares of their uncertainties.}.
These latter two values show a difference within their respective error budgets.
Performing a z-test reveals that the difference of the two values is statistically significant, to at least 2$\sigma$.
Slightly larger velocities at 86\,GHz may indicate that higher frequency observations trace the inner part of the jet sheath, which moves faster, whereas lower frequency 43\,GHz observations detect the slower moving outer part of the sheath \citep[cf. transverse velocity stratification as discussed in, e.g.,][]{Ghisellini05}.
This spine-sheath scenario is more extensively discussed in Sec. \ref{ssec:JColl}.

For earlier epochs (between 1999 and 2002), we track the travelling features further downstream, increasing the number of data points for the fit.
The feature velocities, for which an adequate number of data points for a fit were identified are displayed in Table.~\ref{table:Vel86}.
Among these earlier epoch features, ``W1'' seems to be the most intriguing.
Back-extrapolating to determine the ejection year, we find that the feature emerged in $1981.63\pm2.5$, which corresponds to the onset the total intensity maximum (flare) at cm radio wavelengths (see Fig~\ref{fig:LC}).
``W1'' can be tentatively identified in past maps from the literature.
At 22\,GHz, \cite{Venturi93} reported a travelling feature ejection they called ``C'' in 1986, which can also be identified as a bright feature in the 1\,mas region south-west of the nucleus, in the 43\,GHz map presented in \cite{Romney95}.
The position matches that of feature ``W1'' in our 1999 86\,GHz image.
The trajectory of motion furthermore places ``W1'' in the region where the diffuse blob C2 emerged \citep{Nagai10, Nagai14}.
We included three flux density measurements at 1\,mm from the literature, denoted with green boxes in Fig.~\ref{fig:LC}, to facilitate a comparison between the ejection time of ``W1'' and the flux density peak.
Although there are only three measurements available, the maximum flux at 1\,mm appears roughly in 1982, consistent with the expectation that the flare first became visible at shorter wavelengths.
In both cases, our analysis tentatively connects the appearance of C2 to the major outburst of early 1980s.

For newer epochs, the ejected features seem to move with the expected average velocity (\wavermasyr ~or \waver).
On the other hand, features moving perpendicularly to the bulk jet flow exhibit slower speeds.
A characteristic example is that of ``W10'', which has also been identified by \cite{Oh22}.
The authors find a velocity of $\leq0.03$c, in agreement with our result of \wtenmasyr\ (\wten).
Surprisingly, the emergence of a new feature, ``W14'', in 2019 seems to not follow this rule. 
For it, we find a velocity of \wfourteenmasyr\ (\wfourteen), close to the upper limit of $\sim0.1$c, which has been observed for the core region of \C\ in the past.

The picture gets clearer at 43\,GHz, with just six travelling features needed to describe the central region of \C.
Here we make a comparison with the reported velocity range of $\beta_\textrm{app}\approx(0.086-0.1)$c by \cite{Punsly21}.
The authors of this work used only observations between August 2018 and April 2020 to study the core of \C\ and arrived at this result both by grouping the pair of nuclear CC and by using circular Gaussian component models. 
In this work on the other hand, we did not use all available epochs in that time frame, as the goal was the direct comparison with 86\,GHz images at a comparable cadence. 
We did however double the cadence for 2020 and added the more recent epochs (up to May 2021) to facilitate our analysis.
In our interpretation, three features emerged in that time frame; one feature (``Q4''), which emerged in 2018, one in early 2019 (``Q5''), and one which emerged in early 2020 (``Q6'').
This leads to different feature velocities, with ``Q4'' moving at \qfourmasyr\ (\qfour) and ``Q6'' at \qsixmasyr\ (\qsix).
Averaging them leads to a virtually identical value of \qcombmasyr\ (\qcomb).
The feature ``Q5'' only appears in two epochs, therefore not fulfilling our criterion set for feature velocity determination.
We note here that for this comparison we used the velocity which we computed by fitting only the 43\,GHz epochs.

\cite{Jorstad17} did a similar kinematics analysis for epochs between 2007 and 2013 at 43\,GHz and found moving feature velocities of $\beta_\textrm{app}\sim0.2$c further downstream ($\beta_\textrm{app} = 0.5-3.0$\,mas from the VLBI core) than the region we are studying. 
This is in line with the known acceleration of features downstream along the jet in \C\ \citep{Krichbaum92, Dhawan98, Suzuki12, Punsly21, Hodgson21}.

\subsection{Outburst ejection relations}
\label{ssec:Outburst}

For \C, a conclusive relation between radio-features and $\gamma$-ray flares, as well as their ejection sites, is still elusive \citep[see e.g.][]{Nagai12, Linhoff21, Hodgson18, Hodgson21}.
The emergence of new VLBI features in AGN radio-jets (often but not always) can be associated with the onset of radio-flares \citep[e.g.][]{Savolainen02, Karamanavis16}.
Here, we investigate the possible association between the detected new VLBI features in 3C\,84 and the total intensity light curves, by back-extrapolating their trajectory to estimate the ejection time, which is the time of zero separation from the VLBI core.
We also explore the connection between the flare intensity and ejected feature velocity.

Figure \ref{fig:LC} overlays the ejection time of each feature on the radio and $\gamma$-ray light curves of \C\ in the top panel, with the bottom panel displaying a zoomed in version to aid the reader in comparing the ejection times with the finer details of the variability curves.
The grey shaded area in the top panel of Fig.~\ref{fig:LC} denotes the ejection time of the oldest VLBI feature we can identify in our data set, which is ``W1''. 
Also for the features ''F1'' and ''F2'' only variability data at cm-wavelength are available. 
The flux densities of the cm-light curves are decaying and no correspondence to flare activity is apparent.
However, feature ``F3'' seems to be associated with the onset of a cm-flare peaking in 2013.19. 
At this time no prominent flare in the $\gamma-$ray light curve is obvious.
Feature ``F4'' appears at the onset of a more prominent mm-flare, which is visible in the cm- and in the mm-bands, with a peak in 2016.66.
Interestingly, ``F4'' could be associated with a $\gamma$-ray flare, which occurred at the onset of this radio-flare, and which peaked in 2015.82.
Feature ``F5'' occurred during the decay phase of the cm-/mm-flare of 2016.66, and seems to be corresponding to a rising phase in the $\gamma-$rays, which reached the so far brightest $\gamma-$ray flux peak of \C\ in 2018.35. 
We further note that the peak of this $\gamma-$ray flare occurred {\it after} the decay of the preceding radio-flare, which is quite unusual.
The good correspondence of the ejection times of the features ``F4'' and ``F5'' with two maxima in the $\gamma-$ray light curve, which bracket this prominent cm-/mm-peak, is also remarkable.

The appearance of feature ``F4'' at the onset of the big radio flare peaking in 2016.66 and the local $\gamma$-ray flare during this ejection, suggests that the $\gamma$-ray emission region of this event is located within the VLBI core region.
The relatively long time-lag between the peak of the radio flux in 2016.66 and the peak of the subsequent $\gamma$-ray flare of 2018.35, however, suggests that the $\gamma$-rays of this event are produced further downstream in the jet. 
This hypothesis is further supported by the lack of an enhanced radio flux following this $\gamma$-ray flare. 
It remains unclear, how the the ejection of ``F5'' relates to the observed variability in the radio bands.

\begin{figure}
   \centering
   \includegraphics[scale=0.24]{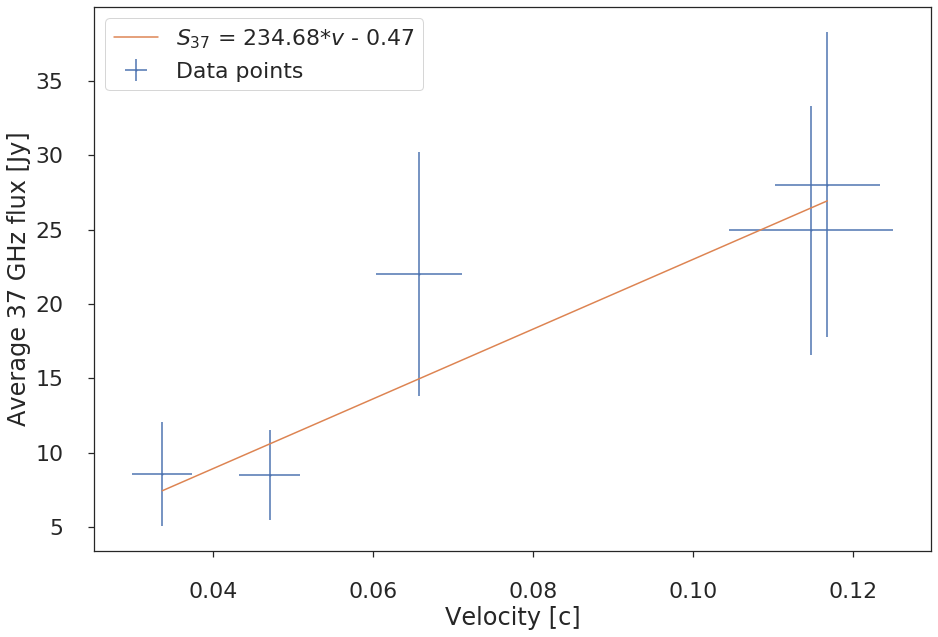}
      \caption{Back-extrapolated, apparent velocity of features ``F1-5'' and the 37\,GHz flux ($S_{37}$) associated with them. 
      A clear, increasing trend is observed.
        }
         \label{fig:f-v}
   \end{figure}

\begin{figure*}
   \centering
   \includegraphics[scale=0.35]{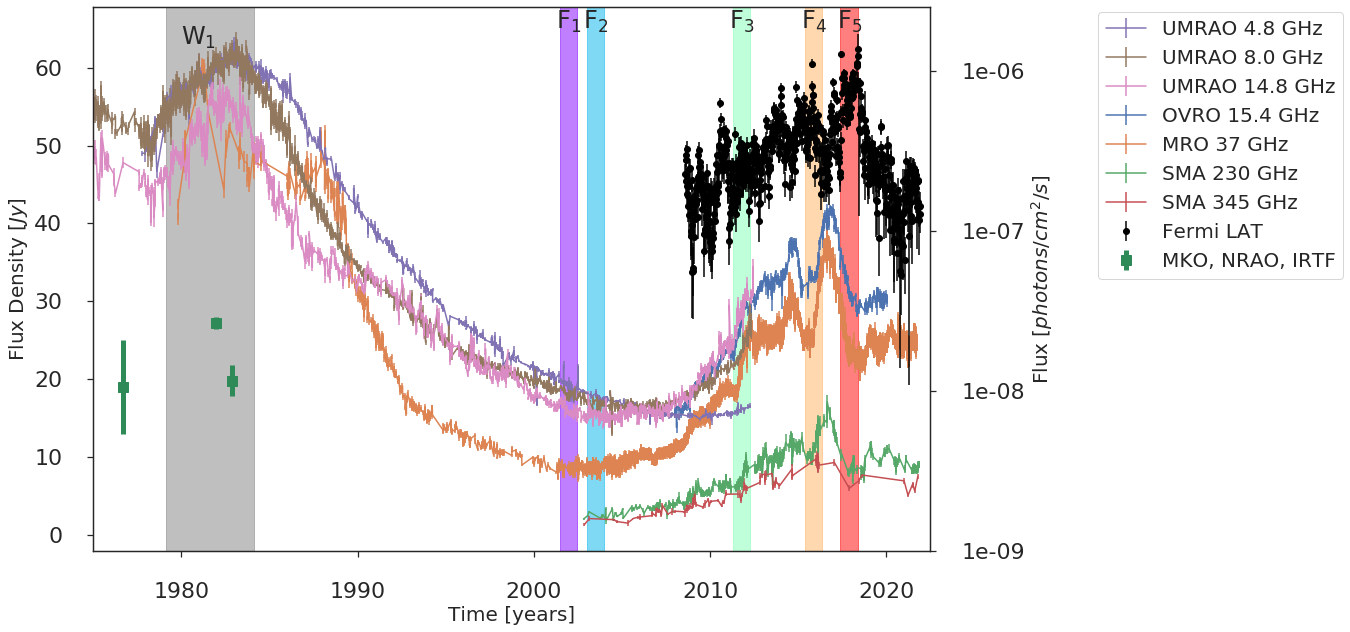}\\
      \includegraphics[scale=0.35]{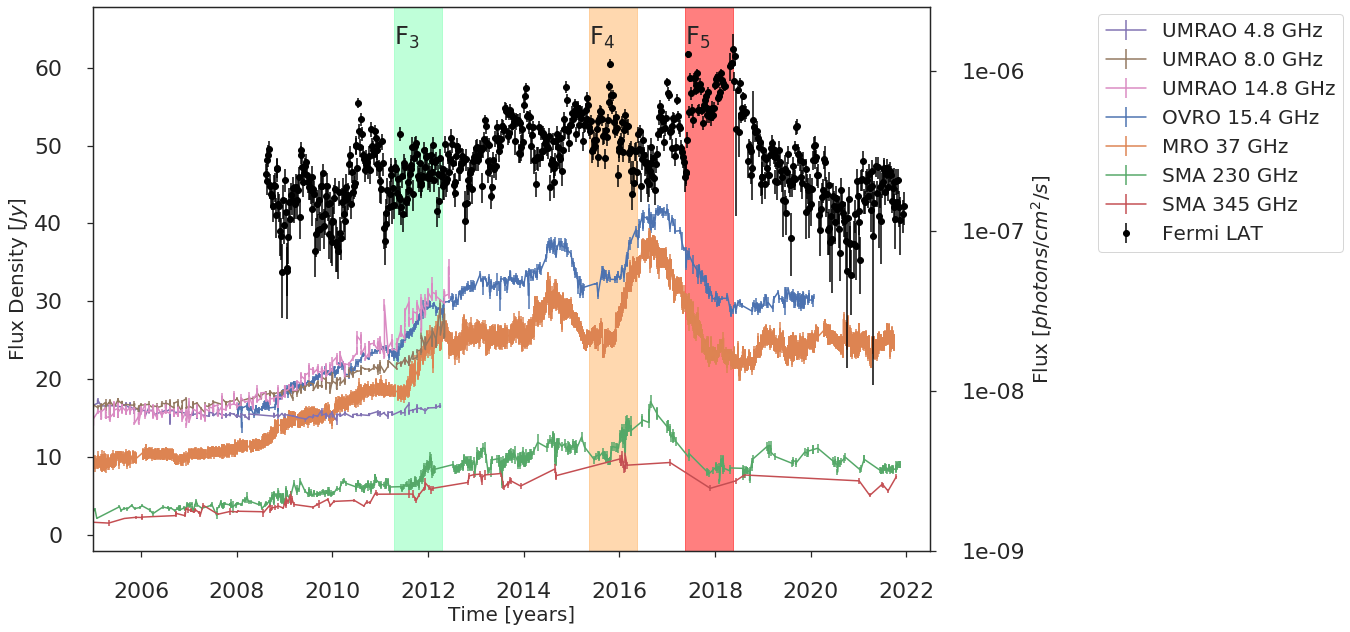}
      \caption{\emph{Top:} Radio light curves of \C\ from 1980 to 2020 at 4.8, 8.0, 14.8, 15, 37, 230, and 345\,GHz (in order of appearance in the legend).
      The black data points correspond to $\gamma$-ray flux.
      The dark-green data points are individual, historical total flux measurements.
      The grey shaded area at the onset of the total intensity peak around 1983 could tentatively be connected to the ejection of feature ``W1'', which may correspond to the prominent C2 region identified in later maps.
      The five other shaded areas are colour-coded in response to the cross-identified features F1 through F5, as described in Fig.~\ref{fig:Vel_Comb}, and denote the approximate ejection time.
      \emph{Bottom:} Zoomed in version between the years 2005 and 2022, of the top panel.
        }
         \label{fig:LC}
   \end{figure*}

Besides the individual associations between the features and the flares, there may also be one between the component velocities and the radio flux densities.
The slowest feature ``F2'' ($\mu^{F_2}\sim0.024$\,mas/yr or $\beta^{F_2}_\textrm{app}\sim0.03c$) seems to correspond to the lowest flux level out of the five features, whereas the fastest feature ``F4'' ($\mu^{F_4}\sim0.09$\,mas/yr or $\beta^{F_4}_\textrm{app}\sim0.11c$) corresponds to the highest flux level. 
The remaining features also follow this pattern (see Fig.~\ref{fig:f-v}). 
This pattern might be transferred also to the $\gamma$-ray flux, with ``F3'', the slower out of the three overlapping features, corresponding to a lower flux level than the two remaining ones. 
Such a behaviour could be explained if jet features move along a curved trajectory, leading to a time variable Doppler beaming.

\subsection{Image stacking}

Even though image stacking smears out details from individual maps, it offers a better insight to the wider jet funnel and into its transverse profile and possible stratification.
For that reason, we averaged all epochs in time at 15, 43 and 86\,GHz and display them in Fig.~\ref{fig:Individual_Stacked}.
The centre of alignment was always chosen to be the core feature (labelled as ``C''), which we defined as the north-westernmost feature in each epoch.
With this choice, we eliminate effects due to time variability of the jet origin, as we can consistently align all images at the same position.
We also refer to a previous core shift analysis \citep{Paraschos21}, which shows a moderate frequency dependence of the VLBI core, typically much smaller than the beam size.
The 86\,GHz image showcases only the nucleus because in the southern region the diffuse flux is mostly resolved out and thus a ridge line cannot be robustly traced.
For completeness, we display the same stacked map of \C\ at 86\,GHz in its entirety, in the bottom right panel of Fig.~\ref{fig:Full86}.
An overview of the stacking, slicing and Gaussian fitting procedure is detailed in Appendix  \ref{App:ImStack}.

\subsection{Spectral index}

Following the procedure described in \cite{Paraschos21}, we created spectral index maps for selected epochs (see Fig.~\ref{fig:SPI}), by performing a two-dimensional cross-correlation analysis.
We used the optical thin and still compact emission region C3 for the alignment, except for the data from October 2020, where C3 was not well determined.
Instead, we used in this observation another optically thin feature, which is located south-west of the VLBI core, and is centred at (0.2, -0.4)\,mas.
In the two earliest epochs, we identify a spectral index gradient along the north-south direction, with spectral indices
between $\alpha^{43-86} \sim3-4$ in the north and $\alpha^{43-86} \sim-2$ in the south.
In September 2014, September 2017 and in October 2020, the orientation of the spectral index gradient has changed towards a northwest-southeast direction.
The spectral indices range again between $\alpha^{43-86}\sim +(2-4)$ in the northwest and $\alpha^{43-86}\sim-2$ in the southeast. Spectral indices much lower than $-2$ are 
likely due to beam resolution effects and/or uncertainties in the flux density
scaling at 86\,GHz and therefore may be not real.
The uncertainty calculation for the individual epochs, as well as the spectral index uncertainty maps (Fig.~\ref{fig:SPI_error}), are summarised in Appendix \ref{App:SPI}.

\section{Discussion}\label{sec:Disc}

\subsection{Projection effects} \label{ssec:Proj}

Our analysis presented in the previous section rests on the assumption, that the jet does not significantly precess and that bright VLBI features are expelled from the VLBI core region, identified as the most luminous feature in the radio maps.
This hypothesis is physically motivated and ties well into the existing literature both for \C\ (c.f. identification scheme in \citealt{Nagai14} and \citealt{Oh22}), as well as for other prominent AGN jets like M\,87 \citep{Kim18} and NGC\,1052 \citep{Baczko16}, to name only a few.

The fact that we are able to cross-identify jet components in 3C\,84 at two frequencies and are able to
describe the motion of the VLBI components by a steady and linear function of time supports the assertion that we are in fact observing features ejected from the central engine. This holds for the feature trajectories from the core up to the $\sim1$\,mas region.
The limited time base of this data set does not allow to see strong deviations from linear motion, but the misalignment between inner (sub-mas) and outer (mas-scale) VLBI jet, suggests
spatially bent and non-ballistical component trajectories, similar to those seen in many
other AGN jets.

\subsection{Jet cross-section} \label{ssec:JXsec}

We used the stacked images to measure the transverse jet profile.
Since the jet of \C\ exhibits an increasingly complex structure with increasing distance from the core (likely due to the contribution of C2 \citealt{Nagai10, Nagai14}), we split the region of interest into two:
the core region ($z_\textrm{0}<0.75$\,mas for 43 and 86\,GHz observations and $z_\textrm{0}<1$\,mas for 15\,GHz ones; here $z_\textrm{0}$ is the distance from the core ($z_\textrm{0}=r|_{x=0}$)), and the region downstream ($z_\textrm{0}>0.75-1$\,mas).
Based on the shape of the intensity slices (see Fig.~\ref{fig:Slices}), we fit a single Gaussian function to the core region of the 86\,GHz image and a double Gaussian function to the downstream region.
For the 15, and 43\,GHz images we fit a double Gaussian function in both regions.
From the single Gaussian fit we determined the full width at half maximum (FWHM) of each slice.
To arrive at the deconvolved jet width, we subtracted the beam from the FWHM in quadrature. 
From the double Gaussian function fit we considered the dominant (brighter) one to measure the transverse width of the main jet.

The results of our analysis are presented in Fig.~\ref{fig:FWHM_fit}.
We find that the measured, deconvolved jet widths of the lower frequencies are overlapping close to the jet origin and the jet width becomes frequency dependent further downstream (see Fig.~\ref{fig:FWHM_fit}), that is, the jet width changes with frequency, within core separations of $1.5 \leq z_\textrm{0} \leq 4$\,mas.
To quantify the degree of change in jet width, we calculate the median of the differences $(w_{15}-w_{43})$ and $(w_{43}-w_{86})$: $\Delta r_{15-43} = \medff$ and $\Delta r_{43-86} = \medfe$.
The fact that the pairwise differences are both positive indicates the presence of a systematic trend.
This is a characteristic property of a stratified jet, where different frequencies could map different parts of the jet \citep{Pelletier89, Sol89, Komissarov90}.
Similar limb brightening in \C\ was reported by \cite{Nagai14, Giovannini18, Kim19}, although these studies were based only on single epochs.
However we note that the 86\,GHz measurements may represent a lower limit to the true jet width, due to beam resolution effects and dynamic range limitations (see also Table~\ref{table:CircBeam}).
In order to test the significance of the smaller 86\,GHz jet width we added a normally distributed random noise signal to the 15 and 43\,GHz stacked images, such that their dynamic range matched that of the 86\,GHz observations. 
Interestingly, we find that the systematic trend persists.
The median values with added noise are $\Delta r^\textrm{n}_{15-43} = \medffnoise$ and $\Delta r^\textrm{n}_{43-86} = \medfenoise$; i. e. still positive.
This indicates that the frequency dependent jet width may be real, even when accounting for dynamic range limitations at 86\,GHz.

\subsection{Jet collimation} \label{ssec:JColl}

A parameter frequently used in discriminating between jet formation models is the collimation profile; i.e. the width of the jet $w$, as a function of the distance from the core $z_\textrm{0}$ \citep[see e.g.][]{Daly88}.
This parameter is fundamentally tied to the characteristics of the jet and the external medium, specifically the ratio of the external pressure $P_{\rm ext}$ to the internal pressure $P_{\text{jet}}$, as has been shown in theoretical models and simulations \citep{Komissarov07, Tchekhovskoy08, Lyubarsky09}.
When the external pressure of the gas is described by a power law of the form $P\propto r^{-p}$, with $p<2$, then the bulk jet flow follows a parabolic profile. On the other hand, if the opposite is true and $p>2$, then the bulk jet flow follows a conical profile \citep{Komissarov09}.
In both cases this pressure is counteracting the force free expansion of the magnetised jet.
In this regime, the power law from Eq. \ref{eq:width} is connected to $P_{\rm ext}$ by $\gamma=p/4$.

By using a stacked image of \C\ (constructed from numerous epochs) to highlight the underlying structure of the jet, we attempt here, for the first time for this source, a comparison of the collimation of the jet profile of \C\ in the nuclear region at three different frequencies.
As shown in Fig.~\ref{fig:FWHM_fit}, we chose to fit a power law function to the initial expansion region where the jet width is increasing but not yet affected by the presence of C2.
The resulting power law indices are $\gamma_{15}=\,$\gammaonefive\ at 15\,GHz, $\gamma_{43}=\,$\gammafourthree\ at 43\,GHz, and $\gamma_{86}=\,$\gammaeightsix\ at 86\,GHz respectively.
These power law index values are summarised in Table~\ref{table:Gamma}.
Our result also appears to align well with the findings of \cite{Oh22}. 
In that paper the authors fit both, a conical and a parabolic jet profile, but could not conclusively select one over the other.
Our measurement of $\gamma_{86}=\,$\gammaeightsix\ at 86\,GHz indicates an initially conical jet profile.

\begin{figure*}
   \centering
   \includegraphics[scale=0.2]{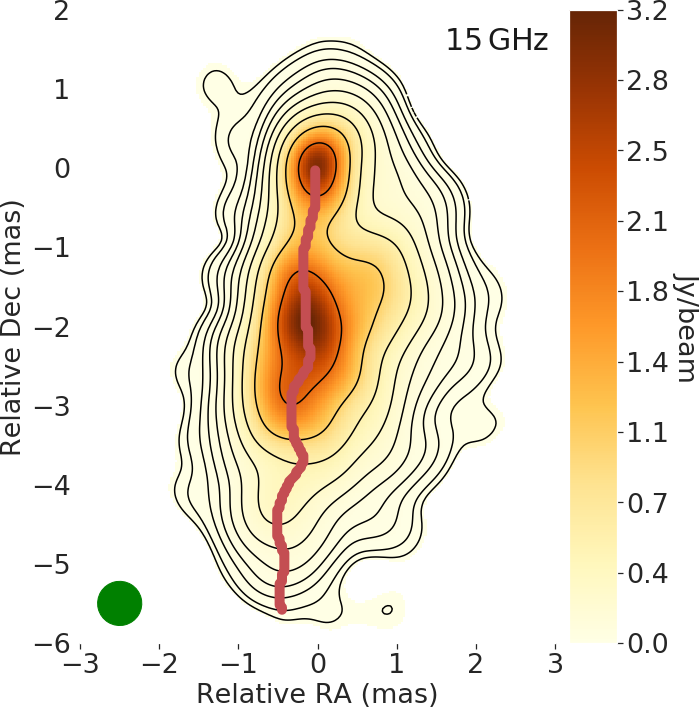}
   \includegraphics[scale=0.2]{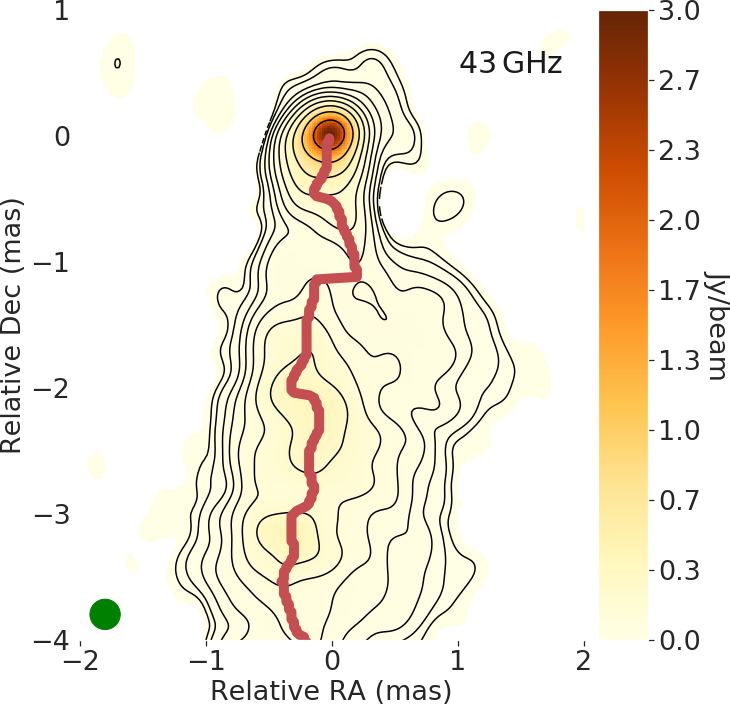}
   \includegraphics[scale=0.2]{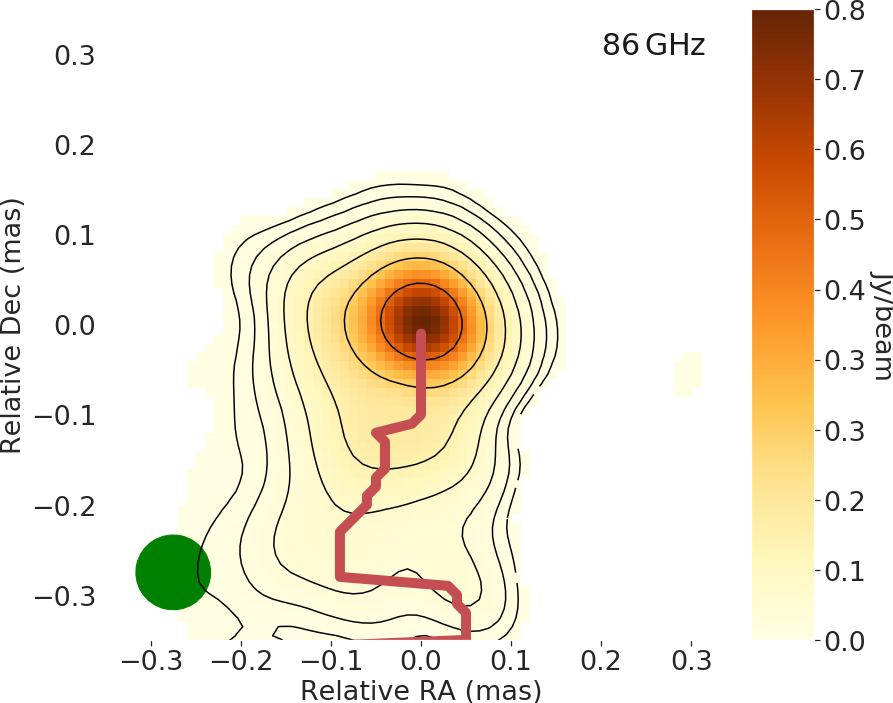}
      \caption{Stacked images of \C\ at 15, 43 and 86 GHz.
      \emph{Left}: 15\,GHz stacked images of epochs from 2010 to 2020, as listed in Table~\ref{table:Beam15}. The green disk represents the common, circular beam used to convolve all images before averaging. The continuous, red line depicts the ridge line, which is the point with the highest intensity at each slice. 
      Each axis is in units of mas. The coloured region and contours correspond to regions with a S/N of at least five.
      Details of the beam size, flux, and rms are listed in Table~\ref{table:CircBeam}.
      The contour levels are -2\%, -1\%, -0.5\%, 0.5\%, 1\%, 2\%, 4\%, 8\%, 16\%, 32\%, and 64\% of the intensity peak.
      \emph{Middle} Same as left panel but for 43\,GHz. An epoch summary is listed in Table~\ref{table:Beam43}. 
      Details of the beam size, flux, and rms are listed in Table~\ref{table:CircBeam}.
      \emph{Right}: Same as left and middle panel but for 86\,GHz. An epoch summary is listed in Table~\ref{table:Beam86}. The common convolving beam is listed in Table~\ref{table:CircBeam}. 
      Details of the beam size, flux, and rms are listed in Table~\ref{table:CircBeam} as well.
      For a zoomed-out view of the 86\,GHz total intensity map at different time bins, including the 2010-2020 bin presented here, refer to Fig.~\ref{fig:Full86}.
    }
         \label{fig:Individual_Stacked}
   \end{figure*}

Figures~\ref{fig:Individual_Stacked} and \ref{fig:Full86} showcase the complex structure of the jet of \C\ in the $z_\textrm{0}>0.75-1$\,mas region.
Plotting the jet width at all three frequencies produces a jagged outline, as a result of both the jet collimation and expansion \citep{Daly88}, as well as image artefacts due to low SNR at the edge of the jet.
On the contrary, the difference between the jet widths at 15, 43, and 86\,GHz provides a better method of investigating and illustrating if a systematic trend is present (see Fig.~\ref{fig:FWHM_fit}, right).
We find a decrease in jet width with increasing frequency, 
suggesting a transverse jet stratification scenario \citep[i.e. spine-sheath structure; see e.g.][]{Pelletier89, Sol89, Komissarov90}. 
Similar stratified jet geometries - also seen via polarisation and rotation measure analyses \citep[e.g.][]{OSullivan09, Hovatta12, Gabuzda17} in other AGN -- could result from twisted magnetic flux tubes \citep[e.g.][]{Tchekhovskoy15},
or hybrid models, where the central black hole and the inner accretion disk both contribute to the jet formation \citep{Paraschos21, Mizuno22}.

Another possible interpretation, however, is that the observed synchrotron radiation reflects magnetised jet plasma that is undergoing rapid synchrotron cooling transverse to the jet flow. Synchrotron losses within the jet might manifest as a frequency dependent jet width, reflecting the radial dependence of the magnetic field strength and the underlying electron power-law distribution across the jet.

A halo/cocoon enveloping the extended emission in the pc-scale region \citep[detected at lower frequencies, e.g.][]{Taylor96, Silver98, Savolainen21} has been put forth in the past as a possible explanation of the jet collimation in \C\ \citep[e.g.][]{Nagai17, Giovannini18}.
However, this halo/cocoon, is extended and exhibits a steep spectrum. Such low surface brightness temperature features are too faint to be detectable at higher frequencies with the present mm-VLBI array.  

We note that an increased, external pressure around the nucleus of \C\ -- e.g. as provided by such a cocoon/halo -- would help to explain the observed downstream acceleration of moving features.
In this picture, the higher external pressure on the nucleus confines the ejected features into a spatially restricted region, which the features can only traverse slowly.
Further downstream, where the external pressure decreases, the features are able to accelerate more freely.

We compare the power law indices of the jet expansion found in this paper to other studies of \C\ from the literature.
\cite{Giovannini18} found a power law index of $\gamma_{86}\sim0.17$, utilising {\it{RadioAstron}} observations of \C\ at 22\,GHz, and \cite{Nagai14} computed a power law index of $\gamma_{86}\sim0.25$ using VLBA observations.
Both these publications use a single epoch and take into account the entirety of the jet in their fitting.
To contrast our results with these publications, we followed a similar procedure, performing a fit over the full length of the jet, which produces $\gamma_{15}^{\text{all}}=\,$\Gammaonefive\ at 15\,GHz, $\gamma_{43}^{\text{all}}=\,$\Gammafourthree\ at 43\,GHz, and $\gamma_{86}^{\text{all}}=\,$\Gammaeightsix\ at 86\,GHz respectively.
Our 43\,GHz estimate of the power law index is very similar to the one from \cite{Nagai14} and the 22\,GHz power law index by \cite{Giovannini18} fits well into the picture of increasing power law indices with frequency for the entirety of the southern bulk jet flow.

Collimation profile studies are also published for several other AGN.
\cite{Boccardi16} found that the power law index of the collimation profile of \object{Cygnus\,A} is $\gamma_{43}\sim0.55$ and \cite{Hada13} showed that for \object{M\,87} the power law index of the collimation profile is $\gamma_\textrm{combined}\sim0.56-0.76$.
While the 86\,GHz data seem to agree with these values, the lower frequency measurements are divergent.
This might indicate source-dependent, intrinsic differences, for example of the jet-power or in the medium which surrounds these jets \citep{Boccardi21}.

\subsection{Implications of the spectral index values and orientation} \label{ssec:JSPI}

The spectral index of a synchrotron self-absorbed emission region usually ranges between $\alpha\leq +2.5$ in the optically thick regime and $-0.5 \leq \alpha \leq -1$ in the optically thin regime (see e.g. \citealt{Rybicki79}). 
An optically thick external absorber located between the source and observer, however 
could alter the spectrum, causing a stronger spectral inversion.

\begin{figure*}
   \centering
   \includegraphics[scale=0.4]{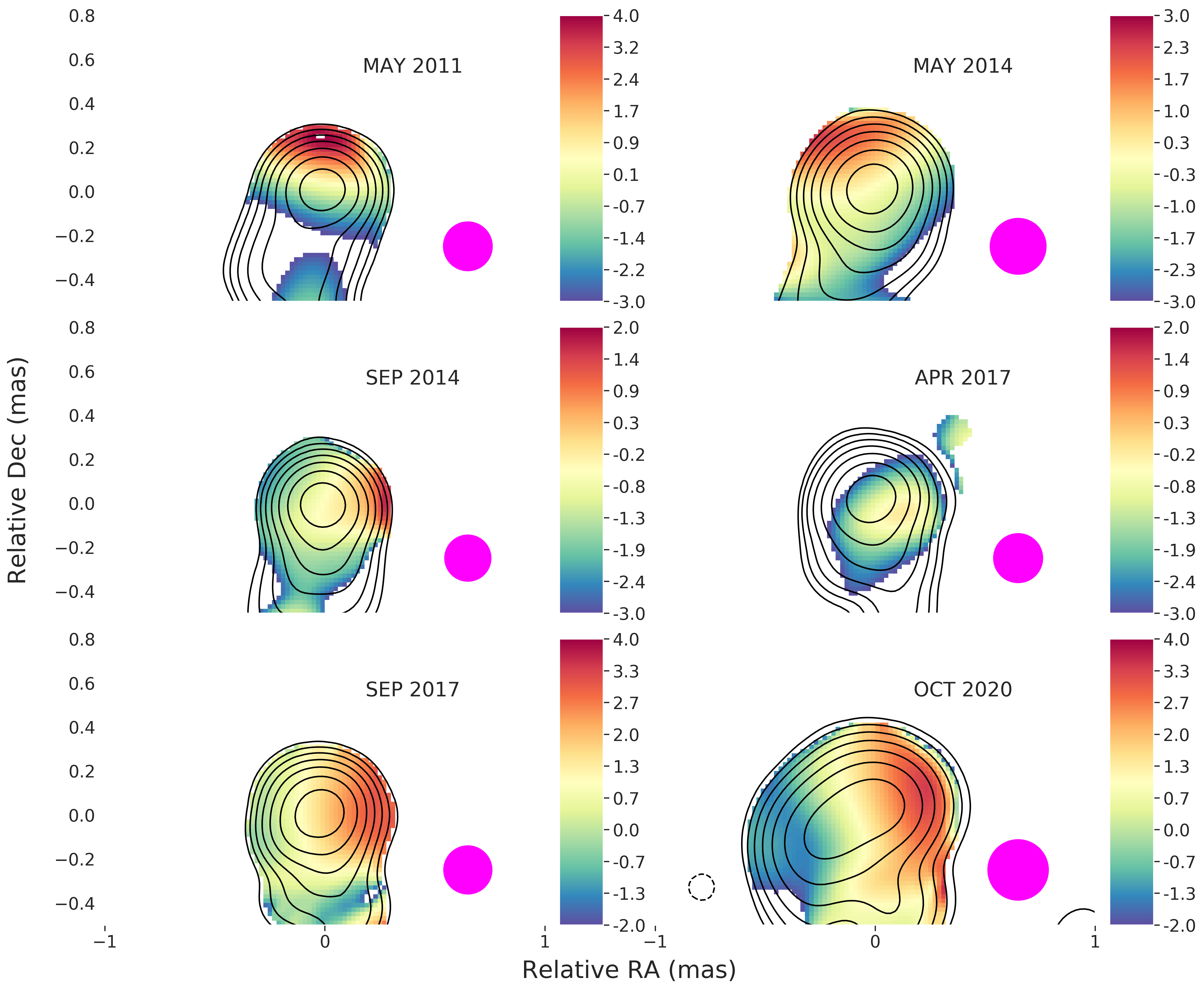}
      \caption{
      Spectral index map of selected epochs.
      The contour levels were set at -1\%, 1\%, 2\%, 4\%, 8\%, 16\%, 32\%, and 64\% of the intensity peak at each epoch.
      The total intensity cutoff was set at $50\,\sigma$.
      The limiting values for the spectral indices are indicated in the colour-bars beside each panel.
      A summary of the image parameters is presented in Table~\ref{table:SPI}.
      The corresponding spectral index uncertainty maps are presented in Fig. \ref{fig:SPI_error}.
      }
         \label{fig:SPI}
   \end{figure*}

\begin{figure*}
   \centering
   \includegraphics[scale=0.3]{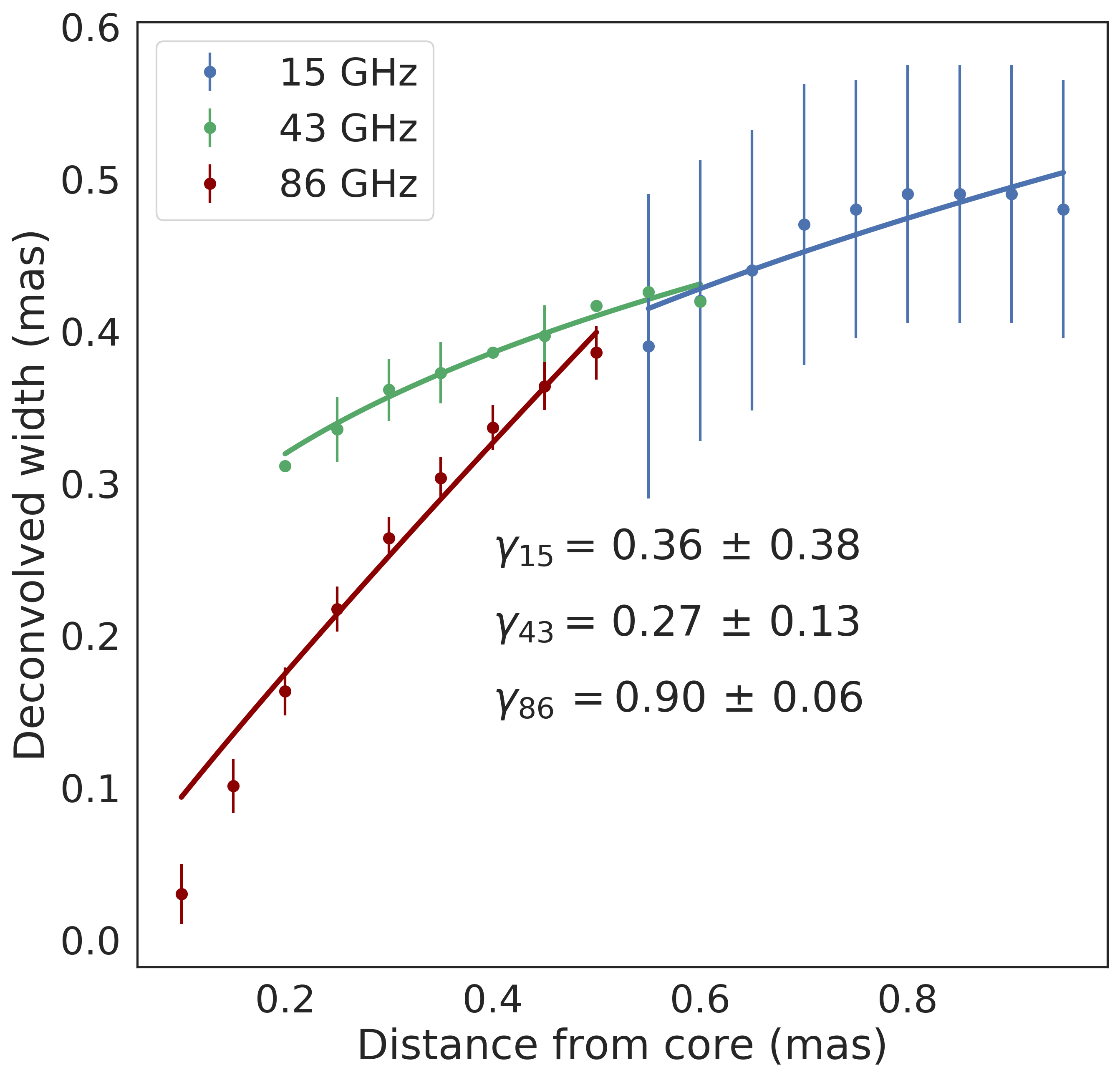}
   \includegraphics[scale=0.3]{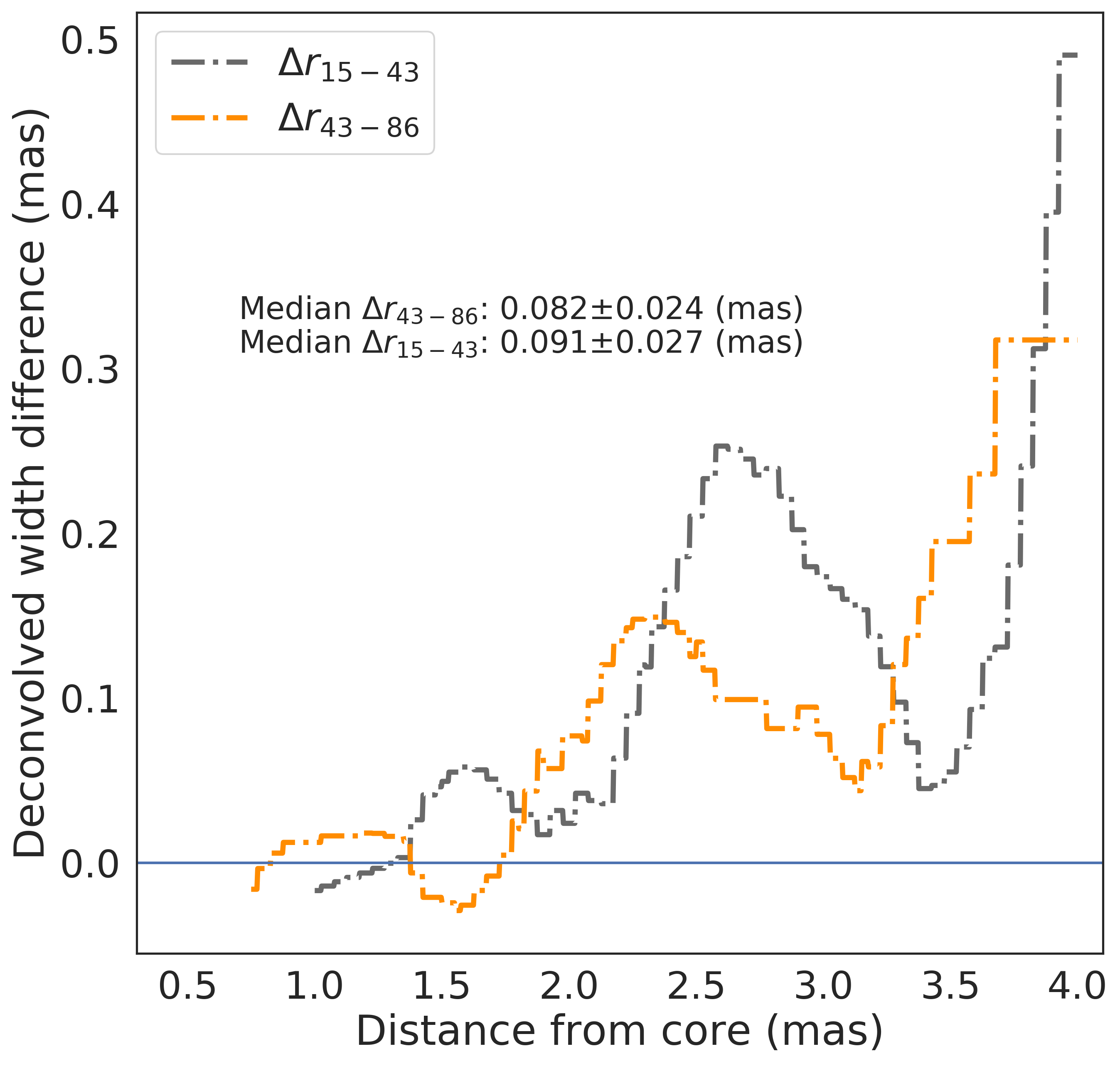}
      \caption{\emph{Left:}  Deconvolved jet width as a function of the distance (projected) from the VLBI core at 15, 43, and 86\,GHz. 
      The blue, green, and red markers correspond to slices taken from the 15, 43, and 86\,GHz images shown in Figs.~\ref{fig:Individual_Stacked} and \ref{fig:Full86} respectively. The corresponding solid lines mark the power law fits with their indices: $\gamma_{15}=\,$\gammaonefive, $\gamma_{43}=\,$\gammafourthree, and $\gamma_{86}=\,$\gammaeightsix\ (see Table~\ref{table:Gamma}).
      All slices are drawn parallel to the abscissa and are evenly spaced at 50\,$\mu$as apart.
      We note that the 86\,GHz data points might be indicating a lower limit for the jet width, due to dynamic range limitations.
      \emph{Right:} Difference between the jet width at 15\,GHz and 43\,GHz ($\Delta r_{15-43}$; dashed grey line) and 43\,GHz and 86\,GHz ($\Delta r_{43-86}$; dashed orange line).
      In both cases, the median values $\Delta r_{15-43}$ and $\Delta r_{43-86}$ are positive.
      }
         \label{fig:FWHM_fit}
   \end{figure*}

The highly inverted spectrum at the northern end of the 3C\,84 jet (see Fig.~\ref{fig:SPI}) therefore suggests free-free absorption from a foreground absorber. In 3C\,84 it could result
for example from a circum-nuclear disk or torus \citep{Walker00}. This view is further supported from recent VLBI observations at 43\,GHz (KaVA) \& 86\,GHz (KVN), though at a lower angular resolution \citep{Wajima20}. A foreground absorber would also
act as an external Faraday screen, explaining the observed frequency dependent polarisation in the central region \citep{Kim19}, and would also explain the apparent absence of the counter-jet on sub-mas scales at 86\,GHz.

The observed variation in die orientation of the spectral index gradient on timescales of
a few years (see Fig.~\ref{fig:SPI}) could challenge
the interpretation via foreground absorption, unless the absorber would be time variable and/or inhomogeneous.
In any case, we have to consider the inclination of the jet relative to the observer.
If we are observing the source as displayed in Fig.~14 in \cite{Kim19}, we would be peering into the bulk jet flow at an angle.
The trajectories of the ejected features could be aligned/misaligned with the line of sight from time to time, thus temporarily affecting the orientation of the spectral index gradient.
In this case the variation of the orientation of the spectral index gradient in the core region could then be explained by the bent jet kinematics (helical motion, rotating jet base) rather than by some small scale variations (sub-parsec) in the foreground screen.
Future monitoring observations and more detailed simulations taking into account time dependent jet bending (helical motion) and absorber variations may shed more light on the observed variation in the spectral index gradient orientation.

\section{Conclusions} \label{sec:Conclusions}

In this work we presented a detailed kinematics analysis of \C\ at 43 and 86\,GHz.
We furthermore established an estimate for the environment around the jet flow in the core region.
Our major findings and conclusions can be summarised as follows:

\begin{enumerate}
    \item We cross-identify moving components ejected from the VLBI core at 43 and 86\,GHz. The components move with velocities of \comavermasyr\ (\comaver) on average. 
    We find marginal evidence ($\sim 2 \sigma$) for a faster motion at 86\,GHz. 
    \item
    The ejection of feature ``W1'', identified at 86\,GHz and possibly corresponding to C2, appears to be temporally coincident with the onset of the total intensity maximum (flare) at cm wavelengths in the early 1980s.
    For the cross-identified radio features (``F1'' through ``F5''), however, a simple one-to-one correspondence between the onset of cm-, mm-, and/or $\gamma$-ray flares seems to be lacking. For components ``F4'' and ``F5'' the data suggest an association between component ejection and a subsequent brightening in the $\gamma-$ray band. The delayed $\gamma-$ray flare peak suggests that the region of $\gamma-$ray production is located not in the VLBI core, but further downstream.    
    
    \item The complexity of the source structure of \C\ complicates the measurement of the transverse jet width. 
    However, in the innermost jet region ($z_\textrm{0} \leq 1$\, mas), we can characterise the transverse jet profile at 86\,GHz by one Gaussian function.
    Further downstream and at 15 \& 43\,GHz, a double Gaussian profile is more appropriate.
    In this downstream region ($z_\textrm{0}>1.5$\,mas) we find evidence for a systematic variation of the jet width with frequency. 
    From a comparison of the brighter Gaussian profile at each frequency, we formally obtain the following power law indices for the transverse jet profiles: at 15\,GHz $\gamma_{15}=\,$\gammaonefive, at 43\,GHz $\gamma_{43}=\,$\gammafourthree, and at 86\,GHz $\gamma_{86}=\,$\gammaeightsix.
    \item The 86\,GHz observations seem to trace the inner sheath part, and suggest a parabolic jet flow profile.
    The 15 and 43\,GHz observations on the other hand seem to outline the outer sheath, characterised by a more conical jet flow profile.
    The different profiles suggest different jet opening angles at each observed frequency and/or effects of synchrotron cooling transverse to the jet axis.
    \item 
    The spectral index images of the individual epochs, reveal a strong spectral index gradient with a time variable orientation. 
    Its position angle may be affected by moving jet features being ejected in different directions. 
    Density or opacity variations in a fore-ground absorber, however, cannot be ruled out.
\end{enumerate}

Overall, our study suggests that the jet stratification could partially explain the motion of the different features seen at 43 and 86\,GHz. Currently, only a handful of VLBI observations of \C\ are available at intermediate frequencies (e.g. 22\,GHz) and none at all at higher frequencies (e.g. 230\,GHz). It is therefore important to bridge this gap by employing more and denser in time sampled VLBI monitoring observations in the near future.
VLBI imaging at even shorter wavelength (eg. with the EHT) will provide an
even higher angular resolution, facilitating more detailed studies of the central SMBH and the matter surrounding it.

\begin{acknowledgements}
      We thank T. Savolainen for providing software to calculate two dimensional cross-correlations.
      We also thank N.~R. MacDonald for the proofreading and fruitful discussions which helped improve this manuscript.
      We thank the anonymous referee for the useful comments.
      G. F. Paraschos is supported for this research by the International Max-Planck Research School (IMPRS) for Astronomy and Astrophysics at the University of Bonn and Cologne. 
      J.-Y. Kim acknowledges support from the National Research Foundation (NRF) of Korea (grant no. 2022R1C1C1005255).
      This research has made use of data obtained with the Global Millimeter VLBI Array (GMVA), which consists of telescopes operated by the MPIfR, IRAM, Onsala, Mets\"ahovi, Yebes, the Korean VLBI Network, the Green Bank Observatory and the Very Long Baseline Array (VLBA). The VLBA and the GBT are a facility of the National Science Foundation operated under cooperative agreement by Associated Universities, Inc. The data were correlated at the correlator of the MPIfR in Bonn, Germany.
      This work makes use of 37 GHz, and 230 and 345\,GHz light curves kindly provided by the Mets\"ahovi Radio Observatory and the Submillimeter Array (SMA), respectively.
      The SMA is a joint project between the Smithsonian Astrophysical Observatory and the Academia Sinica Institute of Astronomy and Astrophysics and is funded by the Smithsonian Institution and the Academia Sinica.
      This research has made use of data from the University of Michigan Radio Astronomy Observatory which has been supported by the University of Michigan and by a series of grants from the National Science Foundation, most recently AST-0607523.
      This work makes use of the Swinburne University of Technology software correlator, developed as part of the Australian Major National Research Facilities Programme and operated under licence. This study makes use of 43\,GHz VLBA data from the VLBA-BU Blazar Monitoring Program (VLBA-BU-BLAZAR; \url{http://www.bu.edu/blazars/VLBAproject.html}), funded by NASA through the Fermi Guest Investigator Program. This research has made use of data from the MOJAVE database that is maintained by the MOJAVE team \citep{Lister09}. This research has made use of the NASA/IPAC Extragalactic Database (NED), which is operated by the Jet Propulsion Laboratory, California Institute of Technology, under contract with the National Aeronautics and Space Administration. This research has also made use of NASA's Astrophysics Data System Bibliographic Services.
      This research has also made use of data from the OVRO 40-m monitoring program \citep{Richards11}, supported by private funding from the California Insitute of Technology and the Max Planck Institute for Radio Astronomy, and by NASA grants NNX08AW31G, NNX11A043G, and NNX14AQ89G and NSF grants AST-0808050 and AST-1109911.
      S.K. acknowledges support from the European Research Council (ERC) under the European Unions Horizon 2020 research and innovation programme under grant agreement No.~771282.
      Finally, this research made use of the following python packages: {\it numpy} \citep{Harris20}, {\it scipy} \citep{2020SciPy-NMeth}, {\it matplotlib} \citep{Hunter07}, {\it astropy} \citep{2013A&A...558A..33A, 2018AJ....156..123A}, {\it pandas} \citep{reback2020pandas, mckinney-proc-scipy-2010}, {\it seaborn} \citep{Waskom21}, and {\it Uncertainties: a Python package for calculations with uncertainties}.
\end{acknowledgements}

\bibliographystyle{aa}
\bibliography{sources}

\begin{appendix}
\section{43 and 86\,GHz contour images}\label{App:Contours}

Here we briefly describe the CC images of \C\ at 86 and 43\,GHz.
They are displayed in Figs. \ref{fig:86mod} and \ref{fig:43mod} respectively. 
Super-imposed are circular Gaussian components used to model the flux.
The features belonging to close in time epochs exhibit similar structure, which confirms our modelling.
The naming convention we used is based on the observing band, plus the timestamp of the ejected feature.
For example, the two features in the top left panel of Fig.~\ref{fig:86mod}, are labelled ``W2'' and ``W1'', because 86\,GHz observations correspond to the W band, and ``W2'' is upstream of ``W1'' and therefore it was ejected first.
All described motions are with regard to the alignment centre, which is the north-westernmost feature in the VLBI core.

\FloatBarrier

\section{Individual feature velocity modelling}\label{App:Comp_vel_mod}  

\begin{figure*}
   \centering
   \includegraphics[scale=0.3]{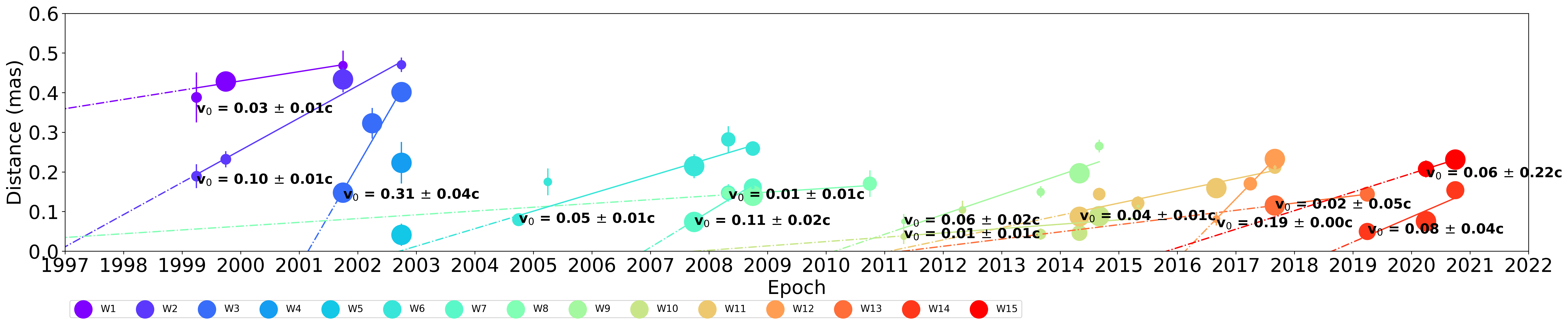}\\
   \includegraphics[scale=0.3]{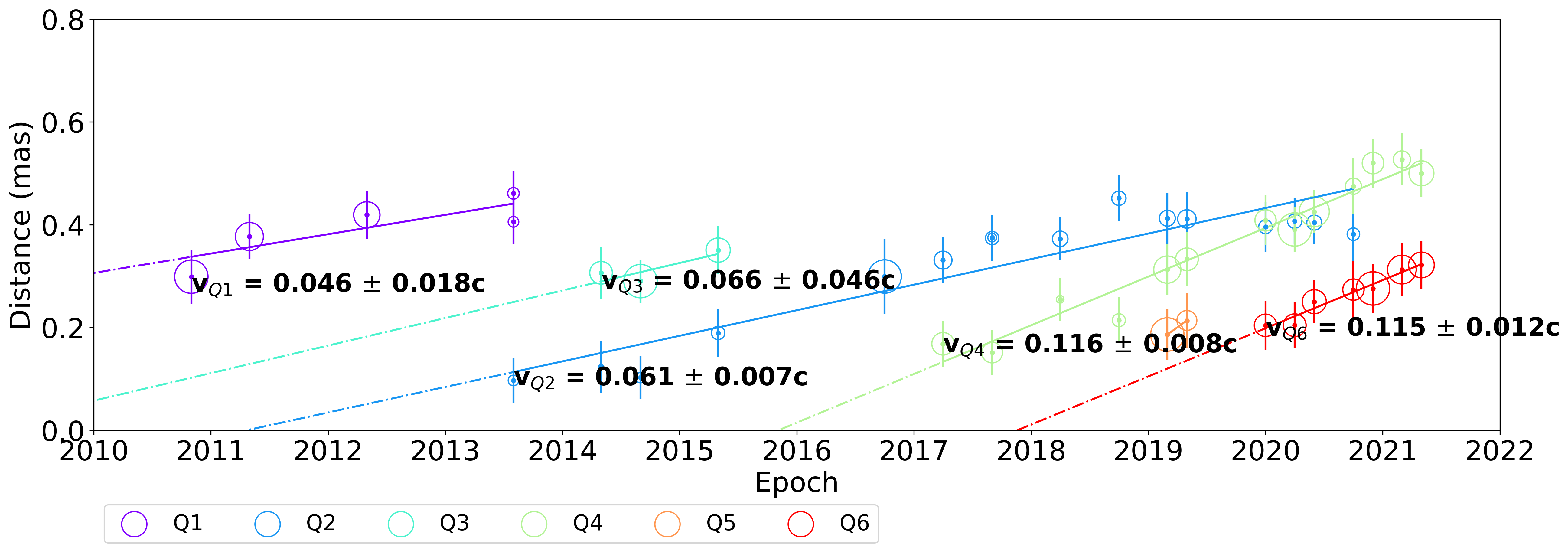}
      \caption{Plot of the motion of all, colour-coded features.
      \emph{Top}: Motion of the 86\,GHz features identified through all available epochs. 
      The velocity fit with the uncertainty is displayed above the first occurrence of each feature.
      The size of each feature is normalised to the flux of the central feature (``C'') of that epoch.
      The velocities are also given in Table~\ref{table:Vel86}.
      \emph{Bottom}: Same as above, but for the 43\,GHz features.
      The velocities are also given in Table~\ref{table:Vel43}.}
         \label{fig:TD4386}
   \end{figure*}

After having identified the travelling features, we used a linear fit of the form: 
\begin{equation}
    d(t) = v_0t+d_0, \label{eq:linear}
\end{equation}
where $d(t)$ is the distance travelled during $t$, $v_0$ is the approximately constant velocity and $d_0$ the initial point of ejection, in order to estimate the velocities.
Since the cross-identified features correspond to the same physical feature we fit a single line to both 43 and 86\,GHz data points, displayed in Fig.~\ref{fig:Vel_Comb}.
The fit parameters are summarised in Table~\ref{table:Vel4386}.
Here me implicitly assumed that position deviations due to opacity shifts are negligible.
We also fit lines to the features individually per frequency.
The fit was applied to the averaged feature positions, calculated from the circular Gaussian components and the CC clusters.
These individual fit parameters are summarised in Tables \ref{table:Vel86}, and \ref{table:Vel43} as well as displayed in Fig.~\ref{fig:TD4386}.
We used the inverse of the uncertainties squared as weights.
For our conservative uncertainty estimation, we used the approach described by the Eqs. (14-5) in \cite{Fomalont99} for the weights, or 1/5th of the beam size (commonly used as a lower limit estimate for sufficiently high SNR images as is the case here; see e.g. \citealt{Hodgson18, Oh22}), depending on which value was greater, for the component distance and component size uncertainties.
For the P. A. errors, we used the simple error propagation of the formula $\theta=2\cdot\arctan \left.( \text{Size}/\text{Distance}\right.)$.
All the aforementioned parameters, for all components, at both 43 and 86\,GHz, are shown in Tables \ref{table:Par43} and \ref{table:Par86}.
With this careful approach we incorporated systemic uncertainties, while also exceeding stochastic noise uncertainties.    
All parameters, describing the individual features, are summed up in Tables \ref{table:Par43} and \ref{table:Par86}.
The velocities together with their uncertainties derived from the fit are listed in Tables \ref{table:Vel86} and \ref{table:Vel43}.

\FloatBarrier

\section{Image stacking}\label{App:ImStack}

Below we present the image stacking and Gaussian function fitting procedure used to create Figs. \ref{fig:Individual_Stacked} and \ref{fig:FWHM_fit}, in detail.
Since individual VLBI images can be subjective and single epochs can suffer from sidelobes and similar related imaging issues, stacking all available epochs and analysing the time averaged image provides the highest level of confidence for our results\footnote{We considered all available epochs up to October 2020 for the 43\,GHz observations, to match the time frame imposed by the 86\,GHz data availability.}.
Before stacking the maps, we convolved each input image with a circular Gaussian beam, the size of which was computed by averaging the major and minor axes of all input images and then calculating the geometrical beam of these averaged major and minor axes.
The values of the convolving beam radii are listed in Table~\ref{table:CircBeam}.
Since all our maps were already aligned at the origin, we need not apply any further shifts, prior to averaging the images.
The time averaged maps are shown in the panels of Fig.~\ref{fig:Individual_Stacked} with details of the contour levels and cut-offs described in the caption.
We implemented three different weighting methods to ascertain which features of the images are robust.
Specifically we applied dynamical range and rms weighting of each epoch, as well as equal weighting for all epochs.
The displayed images in Fig.~\ref{fig:Individual_Stacked} were produced with the latter method.
We find that the lower frequency flux consistently envelops the higher frequency flux.

For the jet width calculation, we make slices oriented parallel to the abscissa.
We choose this approach to focus on the overall bulk jet flow and disregard individual features, which cause the ridge line to deviate locally from the overall north-south flow direction.
To counteract the local deviations and better compare the mean jet widths at each frequency, we performed a Savitzky-Golay filtering \citep{Savitzky64} and interpolation of the data.
Here we define the ridge line as the point of maximum intensity at each slice.
The minimum distance of the first slice intersection with the ridge line to the core is dictated by the beam size; anything smaller we treat as unresolved.

\begin{figure*}
   \centering
   \includegraphics[scale=0.4]{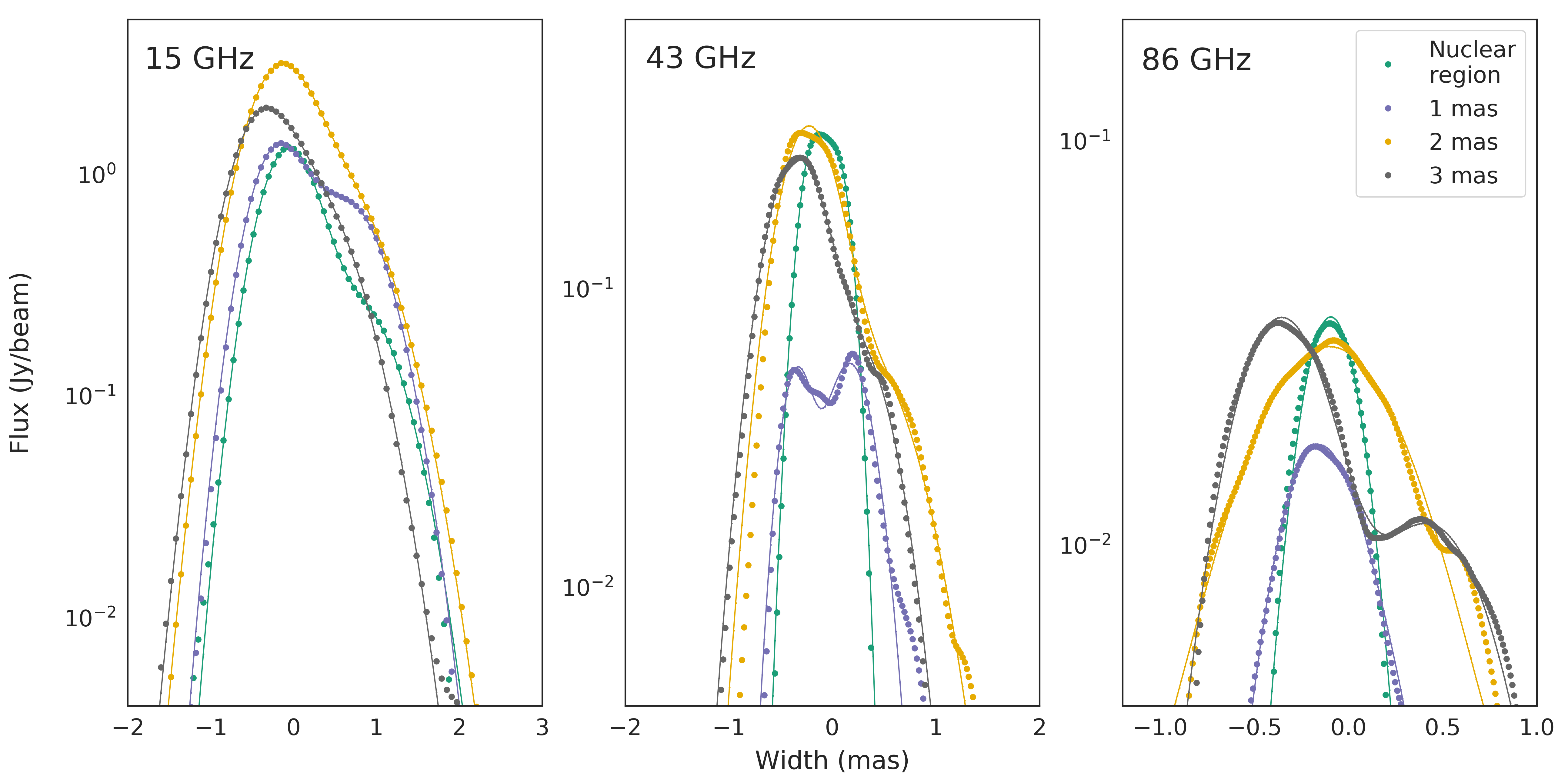}
      \caption{
        Representative transverse jet profiles (dots) and the single and double Gaussian function fits (continuous lines) at 15, 43, and 86\,GHz. 
        Different colours represent slices in the nuclear region ($z_\textrm{0}$\,=\,0.6\,mas at 15\,GHz; $z_\textrm{0}$\,=\,0.4\,mas at 43 and 86\,GHz) and at separations of $z_\textrm{0}$\,=\,1\,mas, 2\,mas, and 3\,mas from the core, with the colour-coding as displayed in the figure legend.
      }
         \label{fig:Slices}
   \end{figure*}

To extract the jet width we use for the core region at 86\,GHz ($z_\textrm{0}<0.75$\,mas) a single Gaussian function of the form:
\begin{equation}
    G_\textrm{tot}(x; \textrm{A}, \mu, \sigma) = \textrm{A} \exp\left.\left(\frac{-(x-\mu)^2}{2\sigma^2}\right.\right) \label{eq:Gauss}
\end{equation}
where $G_\textrm{tot}(x)$ is the Gaussian function at the position $x$, $\textrm{A}$ is the amplitude, $\mu$ is the mean value and $\sigma$ the variance, from which the FWHM of the slices transversely drawn to the bulk jet flow are calculated.
For the 15, and 43\,GHz observations, as well as further downstream for the 86\,GHz observations, we use a sum of two Gaussian functions: $G_\textrm{tot} = G_\textrm{bright}+G_\textrm{faint}$, and considered the FWHM of the brighter Gaussian $G_\textrm{bright}$.
We calculate the deconvolved FWHM denoted as $\theta$, using the following equation:
\begin{equation}
    \theta = \sqrt{{\rm FWHM}^2-{\rm beam}^2}. \label{eq:theta}
\end{equation}
Figure~\ref{fig:Slices} displays the 15, 43, and 86\,GHz jet profile at representative distances from the VLBI core.

For data weighting, we used the rms from the averaged maps.
The uncertainty budget of the fit $\delta\theta$ consists of three parts: (i) the uncertainty of the FWHM ($\delta\text{FWHM}$) of the slice for which we used the \cite{Fomalont99} description; (ii) the uncertainty of the fit ($\delta\text{fit}$); and (iii) an uncertainty introduced by the convolution beam ($\delta\text{beam}$) (assumed to be 1/5th of its radius). 
We added them in quadrature: $\delta\theta=\sqrt{\delta{\rm FWHM}^2+\delta{\rm fit}^2+\delta{\rm beam}^2}$.

We characterise the dependence of the jet width as function of core separation, using a power law of the form:
\begin{equation}
    w(z_\textrm{0}) = c_1z_\textrm{0}^\gamma, \label{eq:width}
\end{equation}
where $w(z_\textrm{0})$ is the width at distance $z_\textrm{0}$ from the core, $c_1$ is a multiplicative constant and $\gamma$ is the power law index describing the jet expansion, with the results showcased in Fig.~\ref{fig:FWHM_fit}.
In Table~\ref{table:Gamma} we summarise the power law index for each frequency.

Having obtained the deconvolved bulk jet flow width, we fit a power law of the form:
\begin{equation}
    w(z_\textrm{0}) = c_1z_\textrm{0}^\gamma, \label{eq:width}
\end{equation}
where $w(z_\textrm{0})$ is the width at distance $z_\textrm{0}$ from the core, $c_1$ is a multiplicative constant and $\gamma$ is the power law index describing the widening rate of the jet, with the results showcased in Fig.~\ref{fig:FWHM_fit}.
The power law index per frequency is displayed in Table~\ref{table:Gamma}.

\FloatBarrier
\section{Spectral index uncertainty} \label{App:SPI}

\begin{figure*}
   \centering
   \includegraphics[scale=0.4]{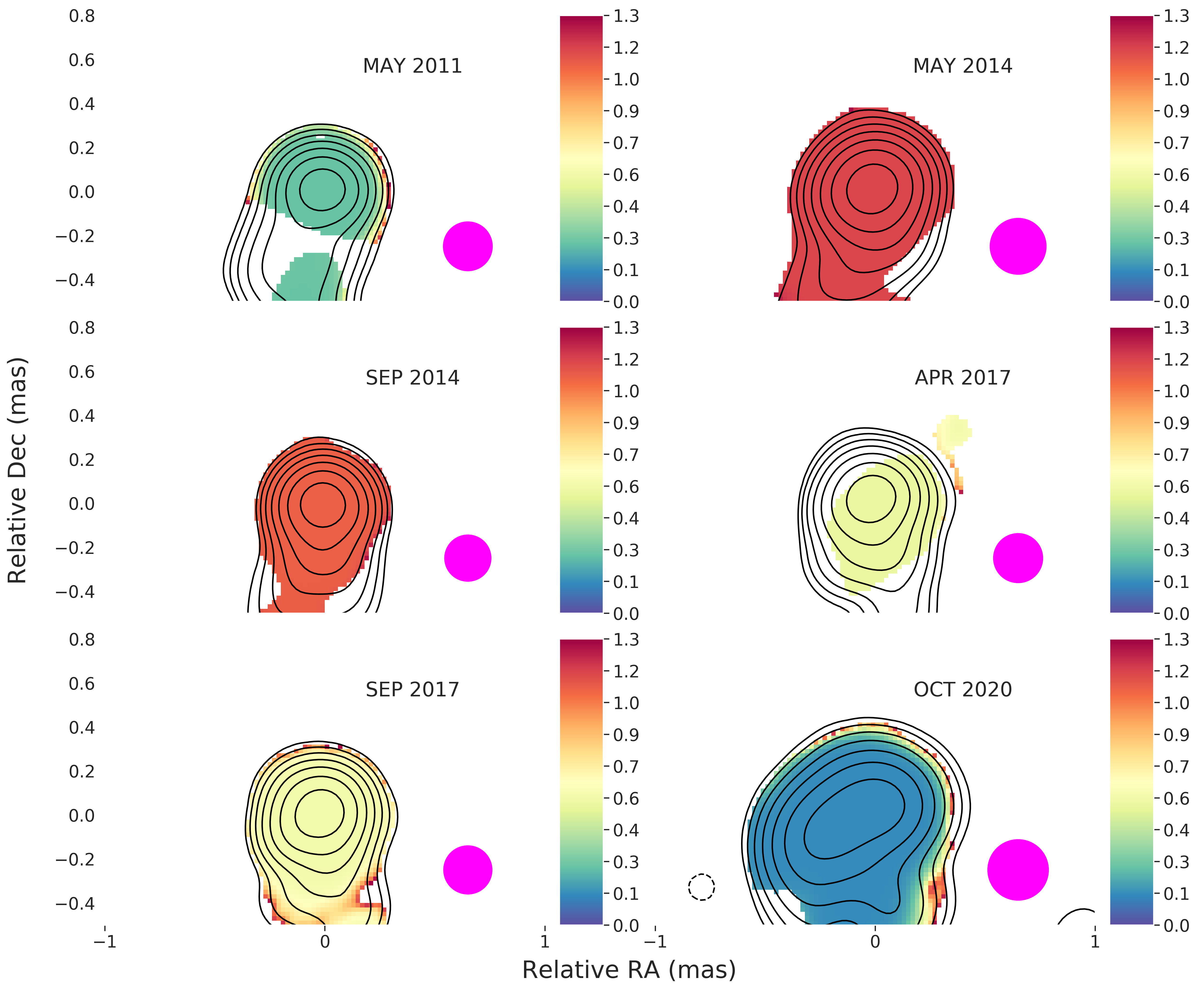}
      \caption{
      Individual spectral index error maps of \C, between 43 and 86\,GHz for the same selected epochs as in Fig.~\ref{fig:SPI}.
      The same contour levels and beam sizes were applied.
      A summary of the image parameters is presented in Table~\ref{table:SPI}.
      The procedure to determine the uncertainties is described in Appendix \ref{App:SPI}.}
         \label{fig:SPI_error}
   \end{figure*}

Figure \ref{fig:SPI_error} showcases the spectral index uncertainty maps of the individual epochs.
The uncertainty in each map can be separated into two parts: the uncertainty in the spectral index gradient position angle, and the uncertainty of the absolute values of the spectral index.
The latter exists because the 86\,GHz VLBI flux measurements are constantly lower than the flux measured with single dish telescopes, as the small VLBI beam resolves out much of the diffuse flux.
Thus the flux at 86\,GHz needs to be scaled up.
Finally, there is some uncertainty in the registration of the phase centre of each epoch (i.e. the location of the core feature ``C'').
\cite{Paraschos21} showed that this uncertainty is less than the pixel scale, and can therefore be neglected.

We follow the procedure described in the appendix of \cite{Kim19} to find the scaling factor $g$ and its uncertainty $\delta g$ per epoch.
The spectral index is defined as:
\begin{equation}
    \alpha = \frac{\log\left(\frac{\rm S_{43}}{g\rm S_{86}} \right)}{\log\left( \frac{\nu_{43}}{\nu_{86}}\right)}, \label{a}
\end{equation}
where ($\nu_{43},\ \rm S_{43}$), and ($\nu_{86},\ \rm S_{86}$) are the frequency and flux at 43 and 86\,GHz respectively.
We then calculated the spectral index uncertainty through standard error propagation, taking into account the uncertainties from the fluxes ($\delta \rm S_{43}$, and $\delta \rm S_{86}$), and from the scaling factor ($\delta g$).
The resulting maps exhibit values between $\alpha^{43-86}\sim0.1-0.7$, except for May 2014 and September 2014 observations, where the uncertainty rises to $\alpha^{43-86}\sim1.0-1.2$.

For the estimation of the robustness of the position angle of the spectral index gradient to the change from north-south to northwest-southeast, we shifted the October 2020 map in both $x$ and $y$ axes and studied the directionality of the spectral index gradient.
In the currently displayed image, the position angle of the shift between the 43 and 86\,GHz image is $\textrm{P. A.} = \sim68^{\circ}$.
We found that in order to produce a north-south oriented spectral index, we needed to shift the same two images by $\textrm{P. A.} = \sim21^{\circ}$, which is a significant offset.
We can thus conclude, that the orientation of the spectral index gradient is robust.

\FloatBarrier
\section{Time binning}\label{App:Bin}

Figure \ref{fig:Full86} displays two different time binning methods of the epochs presented in Fig.~\ref{fig:86mod}. 
The top four panels show a six year bin, whereas the bottom two ones show a $\sim$\,ten year bin.
In the six year binned images, we can recognise the swinging of the P. A. of the southern jet.
In the first bin, the bulk jet flow follow a southwest trajectory. 
In the second bin, we find a purely southern flow.
The third bin is less conclusive, although a tendency towards a southeast trajectory can be recognised, which is then confirmed in the last bin.
We can conclude that the period swing is approximately six, which agrees well with the estimate of five to ten years, found by \cite{Punsly21}.
On the other hand, in the ten year bin, this behaviour is less pronounced.
One conclusion we can draw from the ten year time bins is that the jet has expanded towards the south, from $\sim1.5$\,mas to $\sim3.5$\,mas.

Zooming into the core region in Fig.~\ref{fig:Full86}, the east-west elongation seems absent in the first time bin, and can only be tentatively detected in second time bin.
Then it features more prominently in the third and fourth bins.
Moving further south to the $0.5-1.5$\,mas region, the jet seems ridge brightened, before turning limb brightened in the mid 2000s.
This is in line with the previous work by other authors at lower frequencies (e.g. \citealt{Walker00, Nagai14} at 22 and 43\,GHz).

\FloatBarrier
\section{Parameter tables}\label{App:ParTab}

In this section we present all the tables mentioned in the main text.
Table~\ref{table:Beam86} displays the 86\,GHz epoch, synthesised beam size of the array, pixel scale of the images created, as well as the total intensity maximum in Jy/beam.
Accordingly, Table~\ref{table:Beam43}, and Table~\ref{table:Beam15} refer to the same parameters for the 43 and 15\,GHz epochs.
\cite{Homan02} estimated a flux uncertainty of the order of 5\% to 10\% for individual features in 22\,GHz images, and \cite{Punsly21b} 10\% for 43\,GHz  images. 
Our approach is to be conservative with our error estimation, therefore we factor in a 20\% uncertainty into all our epochs at each frequency.
It should be noted here that the GMVA is known to suffer from large uncertainties in the flux density.
A visibility scaling method is introduced by \cite{Kim19}, culminating in a up-scaling factor of the order two.
Our analysis is independent of the absolute flux densities, thus we opted not to apply this correction.
The reader should therefore keep in mind that the various GMVA fluxes reported in this study might we underestimated by a factor of $\sim2$.

\begin{table*}[h]
\caption{Summary of image parameters at 86\,GHz.}
\label{table:Beam86}
\centering          
\begin{tabular}{cccccc}
Epoch & Beam size (mas) & Beam orientation (deg) & Pixel scale (mas/pixel) & S$_{\text{max}}$ (Jy/beam) & rms$_{\text{mean}}$ (mJy/beam)\\
\hline\hline
 1999.25 &  0.16\,$\times$\,0.07 &  -7.29$^\circ$  &  0.01 &  0.69\,$\pm$\,0.14 &  0.69 \\
 1999.75 &  0.16\,$\times$\,0.06 &  -12.08$^\circ$ &  0.01 &  0.84\,$\pm$\,0.17 &  0.01 \\
 2001.75 &  0.21\,$\times$\,0.04 &  -9.46$^\circ$  &  0.01 &  0.47\,$\pm$\,0.09 &  0.75 \\
 2002.25 &  0.31\,$\times$\,0.06 &  -3.74$^\circ$  &  0.01 &  0.48\,$\pm$\,0.10 &  0.63 \\
 2002.75 &  0.23\,$\times$\,0.04 &  -20.62$^\circ$ &  0.01 &  0.27\,$\pm$\,0.05 &  0.26 \\
 2004.75 &  0.14\,$\times$\,0.04 &  -10.58$^\circ$ &  0.01 &  0.52\,$\pm$\,0.10 &  0.87 \\
 2005.25 &  0.11\,$\times$\,0.04 &  -3.11$^\circ$  &  0.01 &  0.42\,$\pm$\,0.08 &  0.84 \\
 2007.33 &  0.10\,$\times$\,0.04 &  -3.89$^\circ$  &  0.01 &  1.01\,$\pm$\,0.20 &  1.21 \\
 2007.75 &  0.12\,$\times$\,0.04 &  -2.48$^\circ$  &  0.01 &  2.06\,$\pm$\,0.41 &  2.85 \\
 2008.33 &  0.12\,$\times$\,0.04 &  -15.53$^\circ$ &  0.01 &  0.78\,$\pm$\,0.16 &  1.30 \\
 2008.75 &  0.14\,$\times$\,0.05 &  -25.49$^\circ$ &  0.01 &  4.08\,$\pm$\,0.82 &  10.22\\
 2010.75 &  0.09\,$\times$\,0.09 &  0.00$^\circ$   &  0.01 &  0.52\,$\pm$\,0.10 &  0.88 \\
 2011.33 &  0.15\,$\times$\,0.06 &  -16.89$^\circ$ &  0.01 &  0.25\,$\pm$\,0.05 &  0.63 \\
 2012.33 &  0.16\,$\times$\,0.08 &  -2.46$^\circ$  &  0.01 &  0.31\,$\pm$\,0.06 &  0.52 \\
 2013.67 &  0.12\,$\times$\,0.04 &  -11.03$^\circ$ &  0.01 &  0.30\,$\pm$\,0.06 &  0.58 \\
 2014.33 &  0.11\,$\times$\,0.04 &  -11.06$^\circ$ &  0.01 &  1.20\,$\pm$\,0.24 &  0.48 \\
 2014.67 &  0.19\,$\times$\,0.03 &  -20.88$^\circ$ &  0.01 &  0.50\,$\pm$\,0.10 &  1.45 \\
 2015.33 &  0.14\,$\times$\,0.05 &  -13.81$^\circ$ &  0.01 &  0.65\,$\pm$\,0.13 &  0.85 \\
 2016.67 &  0.17\,$\times$\,0.04 &  -21.96$^\circ$ &  0.01 &  0.63\,$\pm$\,0.13 &  0.25 \\
 2017.25 &  0.10\,$\times$\,0.04 &  -13.30$^\circ$ &  0.01 &  0.25\,$\pm$\,0.05 &  0.16 \\
 2017.67 &  0.11\,$\times$\,0.05 &  -24.02$^\circ$ &  0.01 &  2.87\,$\pm$\,0.57 &  9.03 \\
 2019.25 &  0.15\,$\times$\,0.04 &  -29.45$^\circ$ &  0.01 &  0.67\,$\pm$\,0.13 &  1.21 \\
 2020.25 &  0.11\,$\times$\,0.05 &  -16.07$^\circ$ &  0.01 &  0.54\,$\pm$\,0.11 &  3.16 \\
 2020.75 &  0.15\,$\times$\,0.04 &  -7.89$^\circ$  &  0.01 &  0.35\,$\pm$\,0.07 &  0.66 \\
\hline
\end{tabular}
\end{table*}

\begin{table*}[h]
\caption{Summary of image parameters at 43\,GHz.}           
\label{table:Beam43}     
\centering               
\begin{tabular}{cccccc}
Epoch & Beam size (mas) & Beam orientation (deg) & Pixel scale (mas/pixel) & S$_{\text{max}}$ (Jy/beam) & rms$_{\text{mean}}$ (mJy/beam)\\
\hline\hline
 2010.83 &  0.45\,$\times$\,0.15 &  -22.70$^\circ$ &  0.02 &  2.20\,$\pm$\,0.44 &  0.85\\
 2011.33 &  0.35\,$\times$\,0.14 &  8.69$^\circ$   &  0.02 &  1.92\,$\pm$\,0.38 &  0.32\\
 2012.33 &  0.36\,$\times$\,0.15 &  16.47$^\circ$  &  0.02 &  2.13\,$\pm$\,0.43 &  0.58\\
 2013.58 &  0.31\,$\times$\,0.15 &  -3.40$^\circ$  &  0.02 &  1.90\,$\pm$\,0.38 &  0.19\\
 2014.33 &  0.40\,$\times$\,0.16 &  16.26$^\circ$  &  0.02 &  3.43\,$\pm$\,0.69 &  0.28\\
 2014.67 &  0.34\,$\times$\,0.13 &  -1.14$^\circ$  &  0.02 &  3.20\,$\pm$\,0.64 &  0.20\\
 2015.33 &  0.29\,$\times$\,0.20 &  -4.05$^\circ$  &  0.02 &  3.93\,$\pm$\,0.79 &  0.42\\
 2016.75 &  0.52\,$\times$\,0.26 &  23.41$^\circ$  &  0.02 &  7.81\,$\pm$\,1.56 &  2.22\\
 2017.25 &  0.33\,$\times$\,0.15 &  1.75$^\circ$   &  0.02 &  3.47\,$\pm$\,0.69 &  0.72\\
 2017.67 &  0.34\,$\times$\,0.14 &  8.37$^\circ$   &  0.02 &  3.34\,$\pm$\,0.67 &  0.44\\
 2018.25 &  0.28\,$\times$\,0.15 &  -6.95$^\circ$  &  0.02 &  2.47\,$\pm$\,0.49 &  1.02\\
 2018.75 &  0.31\,$\times$\,0.16 &  6.05$^\circ$   &  0.02 &  3.53\,$\pm$\,0.71 &  1.15\\
 2019.16 &  0.35\,$\times$\,0.18 &  7.39$^\circ$   &  0.02 &  2.51\,$\pm$\,0.50 &  0.90\\
 2019.33 &  0.41\,$\times$\,0.17 &  -12.05$^\circ$ &  0.02 &  2.40\,$\pm$\,0.48 &  0.87\\
 2020.00 &  0.35\,$\times$\,0.17 &  -2.33$^\circ$  &  0.02 &  1.73\,$\pm$\,0.35 &  0.30\\
 2020.25 &  0.33\,$\times$\,0.15 &  3.25$^\circ$   &  0.02 &  2.24\,$\pm$\,0.45 &  0.37\\
 2020.42 &  0.28\,$\times$\,0.16 &  11.05$^\circ$  &  0.02 &  1.71\,$\pm$\,0.34 &  0.58\\
 2020.75 &  0.44\,$\times$\,0.17 &  4.61$^\circ$   &  0.02 &  1.81\,$\pm$\,0.36 &  0.40\\
 2020.92 &  0.36\,$\times$\,0.16 &  7.71$^\circ$   &  0.02 &  1.89\,$\pm$\,0.38 &  0.59\\
 2021.16 &  0.36\,$\times$\,0.18 &  12.62$^\circ$  &  0.02 &  1.59\,$\pm$\,0.32 &  1.45\\
 2021.33 &  0.29\,$\times$\,0.19 &  -16.99$^\circ$ &  0.02 &  2.08\,$\pm$\,0.42 &  1.03\\
\hline
\end{tabular}
\end{table*}

\begin{table*}[h]
\caption{Summary of image parameters at 15\,GHz.}             
\label{table:Beam15}    
\centering   
\begin{tabular}{cccccc}
Epoch & Beam size (mas) & Beam orientation (deg) & Pixel scale (mas/pixel) & S$_\text{max}$ (Jy/beam) & rms$_{\text{mean}}$ (mJy/beam)\\
\hline\hline
 2010.58 &  0.69\,$\times$\,0.47 &  -11.98$^\circ$ &  0.03 &  4.9$\,\pm\,$0.98  &  1.64\\
 2011.41 &  0.66\,$\times$\,0.49 &  -12.79$^\circ$ &  0.03 &  4.7$\,\pm\,$0.94  &  1.33\\
 2012.16 &  0.91\,$\times$\,0.56 &  25.77$^\circ$  &  0.03 &  8.99$\,\pm\,$1.8  &  1.34\\
 2013.50 &  0.72\,$\times$\,0.55 &  -57.56$^\circ$ &  0.03 &  7.39$\,\pm\,$1.48 &  0.91\\
 2014.50 &  0.64\,$\times$\,0.38 &  -7.82$^\circ$  &  0.03 &  6.74$\,\pm\,$1.35 &  0.65\\
 2015.41 &  0.64\,$\times$\,0.44 &  -28.44$^\circ$ &  0.03 &  4.05$\,\pm\,$0.81 &  0.96\\
 2016.42 &  0.64\,$\times$\,0.36 &  -7.52$^\circ$  &  0.03 &  8.19$\,\pm\,$1.64 &  0.67\\
 2017.25 &  0.61\,$\times$\,0.38 &  -11.74$^\circ$ &  0.03 &  5.15$\,\pm\,$1.03 &  0.67\\
 2018.50 &  0.7\,$\times$\,0.39  &  -8.97$^\circ$  &  0.03 &  3.74$\,\pm\,$0.75 &  0.97\\
 2019.41 &  0.86\,$\times$\,0.46 &  -24.84$^\circ$ &  0.03 &  5.22$\,\pm\,$1.04 &  0.75\\
 2020.50 &  0.67\,$\times$\,0.37 &  -4.10$^\circ$   &  0.03 &  3.47$\,\pm\,$0.69 &  0.81\\
\hline
\end{tabular}
\end{table*}

Tables \ref{table:Par43} and \ref{table:Par86} summarise all the moving features we identified, at 86 and 43\,GHz respectively.
The first column denotes the individual feature ID.
The second column is the observing epoch; column three denotes the flux of the fit feature.
The final three columns correspond in sequence to the distance and the P. A. between each feature and the core, and the size of the feature, defined as the width of the feature.

\begin{table*}[h]
\caption{Physical parameters of each model feature at 43\,GHz.}   
\label{table:Par43}      
\centering               
\begin{tabular}{cccccc}
ID & Epoch & Flux Density (Jy/beam) & Distance (mas) & P. A. (deg) & Size (mas)\\
\hline\hline
 Q1 &  2010.83 &  2.29\,$\pm$\,0.46 &  0.33\,$\pm$\,0.05 &  151.28\,$\pm$\,13.15  &  0.17\,$\pm$\,0.03\\
 Q1 &  2011.33 &  1.61\,$\pm$\,0.32 &  0.37\,$\pm$\,0.04 &  161.39\,$\pm$\,10.89  &  0.23\,$\pm$\,0.02\\
 Q1 &  2012.33 &  1.42\,$\pm$\,0.28 &  0.42\,$\pm$\,0.05 &  170.78\,$\pm$\,12.28  &  0.35\,$\pm$\,0.02\\
 Q1 &  2013.58 &  0.23\,$\pm$\,0.17 &  0.41\,$\pm$\,0.04 &  -176.01\,$\pm$\,7.29  &  0.13\,$\pm$\,0.02\\
 Q1 &  2013.58 &  0.27\,$\pm$\,0.21 &  0.47\,$\pm$\,0.04 &  154.38\,$\pm$\,5.72   &  0.10\,$\pm$\,0.02\\
 \hline
 Q2 &  2013.58 &  0.25\,$\pm$\,0.05 &  0.10\,$\pm$\,0.04 &  107.37\,$\pm$\,39.05  &  0.06\,$\pm$\,0.02\\
 Q2 &  2014.33 &  0.06\,$\pm$\,0.01 &  0.12\,$\pm$\,0.05 &  141.03\,$\pm$\,37.77  &  0.08\,$\pm$\,0.03\\
 Q2 &  2014.67 &  0.24\,$\pm$\,0.05 &  0.12\,$\pm$\,0.04 &  152.29\,$\pm$\,39.77  &  0.09\,$\pm$\,0.02\\
 Q2 &  2015.33 &  0.41\,$\pm$\,0.08 &  0.19\,$\pm$\,0.05 &  154.31\,$\pm$\,19.64  &  0.09\,$\pm$\,0.02\\
 Q2 &  2016.75 &  2.55\,$\pm$\,1.91 &  0.29\,$\pm$\,0.16 &  -172.28\,$\pm$\,75.31 &  0.32\,$\pm$\,0.08\\
 Q2 &  2017.25 &  0.74\,$\pm$\,0.15 &  0.33\,$\pm$\,0.04 &  -165.23\,$\pm$\,10.39 &  0.15\,$\pm$\,0.02\\
 Q2 &  2017.67 &  0.41\,$\pm$\,0.08 &  0.37\,$\pm$\,0.04 &  -166.32\,$\pm$\,7.82  &  0.11\,$\pm$\,0.02\\
 Q2 &  2017.67 &  0.17\,$\pm$\,0.03 &  0.38\,$\pm$\,0.04 &  160.18\,$\pm$\,9.78   &  0.20\,$\pm$\,0.02\\
 Q2 &  2018.25 &  0.55\,$\pm$\,0.11 &  0.37\,$\pm$\,0.04 &  -170.32\,$\pm$\,10.06 &  0.23\,$\pm$\,0.02\\
 Q2 &  2018.75 &  0.47\,$\pm$\,0.09 &  0.44\,$\pm$\,0.04 &  -161.40\,$\pm$\,7.31  &  0.17\,$\pm$\,0.02\\
 Q2 &  2019.16 &  0.58\,$\pm$\,0.12 &  0.40\,$\pm$\,0.05 &  -170.28\,$\pm$\,10.08 &  0.20\,$\pm$\,0.03\\
 Q2 &  2019.33 &  0.78\,$\pm$\,0.16 &  0.42\,$\pm$\,0.05 &  -172.00\,$\pm$\,11.45 &  0.26\,$\pm$\,0.03\\
 Q2 &  2020.00 &  0.43\,$\pm$\,0.09 &  0.40\,$\pm$\,0.05 &  -176.60\,$\pm$\,9.88  &  0.20\,$\pm$\,0.02\\
 Q2 &  2020.25 &  0.47\,$\pm$\,0.09 &  0.41\,$\pm$\,0.04 &  -174.63\,$\pm$\,6.59  &  0.07\,$\pm$\,0.02\\
 Q2 &  2020.42 &  0.52\,$\pm$\,0.10 &  0.40\,$\pm$\,0.04 &  -177.43\,$\pm$\,6.88  &  0.12\,$\pm$\,0.02\\
 Q2 &  2020.75 &  0.36\,$\pm$\,0.07 &  0.38\,$\pm$\,0.06 &  -178.46\,$\pm$\,11.11 &  0.17\,$\pm$\,0.03\\
 \hline
 Q3 &  2014.33 &  0.50\,$\pm$\,0.10 &  0.31\,$\pm$\,0.05 &  163.50\,$\pm$\,20.34  &  0.29\,$\pm$\,0.03\\
 Q3 &  2014.67 &  1.08\,$\pm$\,0.81 &  0.29\,$\pm$\,0.04 &  177.71\,$\pm$\,12.79  &  0.17\,$\pm$\,0.02\\
 Q3 &  2015.33 &  0.57\,$\pm$\,0.11 &  0.35\,$\pm$\,0.05 &  -162.78\,$\pm$\,11.54 &  0.19\,$\pm$\,0.02\\
 \hline
 Q4 &  2017.25 &  0.68\,$\pm$\,0.14 &  0.17\,$\pm$\,0.04 &  136.26\,$\pm$\,20.97  &  0.08\,$\pm$\,0.02\\
 Q4 &  2017.67 &  0.58\,$\pm$\,0.12 &  0.16\,$\pm$\,0.04 &  125.81\,$\pm$\,21.33  &  0.08\,$\pm$\,0.02\\
 Q4 &  2018.25 &  0.08\,$\pm$\,0.02 &  0.25\,$\pm$\,0.04 &  125.67\,$\pm$\,12.58  &  0.12\,$\pm$\,0.02\\
 Q4 &  2018.75 &  0.23\,$\pm$\,0.05 &  0.21\,$\pm$\,0.04 &  106.07\,$\pm$\,23.73  &  0.18\,$\pm$\,0.02\\
 Q4 &  2019.16 &  0.99\,$\pm$\,0.20 &  0.31\,$\pm$\,0.05 &  138.69\,$\pm$\,12.45  &  0.15\,$\pm$\,0.03\\
 Q4 &  2019.33 &  0.69\,$\pm$\,0.14 &  0.34\,$\pm$\,0.05 &  140.35\,$\pm$\,10.65  &  0.11\,$\pm$\,0.03\\
 Q4 &  2020.00 &  0.60\,$\pm$\,0.12 &  0.41\,$\pm$\,0.05 &  142.80\,$\pm$\,7.26   &  0.09\,$\pm$\,0.02\\
 Q4 &  2020.25 &  1.50\,$\pm$\,1.12 &  0.38\,$\pm$\,0.04 &  143.21\,$\pm$\,8.19   &  0.13\,$\pm$\,0.02\\
 Q4 &  2020.42 &  1.22\,$\pm$\,0.91 &  0.43\,$\pm$\,0.04 &  145.06\,$\pm$\,7.42   &  0.18\,$\pm$\,0.02\\
 Q4 &  2020.75 &  0.35\,$\pm$\,0.07 &  0.49\,$\pm$\,0.06 &  153.65\,$\pm$\,7.26   &  0.13\,$\pm$\,0.03\\
 Q4 &  2020.92 &  0.62\,$\pm$\,0.12 &  0.52\,$\pm$\,0.05 &  131.26\,$\pm$\,5.29   &  0.02\,$\pm$\,0.02\\
 Q4 &  2021.16 &  0.40\,$\pm$\,0.08 &  0.55\,$\pm$\,0.05 &  129.77\,$\pm$\,5.68   &  0.11\,$\pm$\,0.03\\
 Q4 &  2021.33 &  0.83\,$\pm$\,0.17 &  0.50\,$\pm$\,0.05 &  124.54\,$\pm$\,5.37   &  0.03\,$\pm$\,0.02\\
 \hline
 Q5 &  2019.16 &  1.13\,$\pm$\,0.85 &  0.19\,$\pm$\,0.05 &  90.36\,$\pm$\,18.47   &  0.06\,$\pm$\,0.03\\
 Q5 &  2019.33 &  0.40\,$\pm$\,0.08 &  0.21\,$\pm$\,0.05 &  98.63\,$\pm$\,17.98   &  0.09\,$\pm$\,0.03\\
 \hline
 Q6 &  2020.00 &  1.70\,$\pm$\,0.34 &  0.21\,$\pm$\,0.05 &  107.76\,$\pm$\,19.15  &  0.11\,$\pm$\,0.02\\
 Q6 &  2020.25 &  1.76\,$\pm$\,0.35 &  0.21\,$\pm$\,0.04 &  106.71\,$\pm$\,14.60  &  0.07\,$\pm$\,0.02\\
 Q6 &  2020.42 &  1.96\,$\pm$\,0.39 &  0.25\,$\pm$\,0.04 &  111.53\,$\pm$\,12.84  &  0.11\,$\pm$\,0.02\\
 Q6 &  2020.75 &  1.54\,$\pm$\,0.31 &  0.27\,$\pm$\,0.06 &  114.92\,$\pm$\,13.57  &  0.09\,$\pm$\,0.03\\
 Q6 &  2020.92 &  3.78\,$\pm$\,0.76 &  0.28\,$\pm$\,0.05 &  112.53\,$\pm$\,19.62  &  0.24\,$\pm$\,0.02\\
 Q6 &  2021.16 &  2.80\,$\pm$\,0.56 &  0.32\,$\pm$\,0.05 &  113.74\,$\pm$\,16.76  &  0.24\,$\pm$\,0.03\\
 Q6 &  2021.33 &  2.24\,$\pm$\,0.45 &  0.31\,$\pm$\,0.05 &  110.00\,$\pm$\,11.85  &  0.14\,$\pm$\,0.02\\
\hline
\end{tabular}
\end{table*}

 \begin{table*}[h]
\caption{Physical parameters of each model feature at 86\,GHz.}  
\label{table:Par86}
\centering         
\begin{tabular}{cccccc}
ID & Epoch & Flux Density (Jy/beam) & Distance (mas) & P. A. (deg) & Size (mas)\\
\hline\hline
 W1  &  1999.25 &  0.11\,$\pm$\,0.02 &  0.38\,$\pm$\,0.19 &  -163.48\,$\pm$\,39.86 &  0.19\,$\pm$\,0.09\\
 W1  &  1999.75 &  0.42\,$\pm$\,0.08 &  0.42\,$\pm$\,0.07 &  -164.85\,$\pm$\,10.54 &  0.07\,$\pm$\,0.04\\
 W1  &  2001.75 &  0.08\,$\pm$\,0.02 &  0.47\,$\pm$\,0.07 &  -164.83\,$\pm$\,9.04  &  0.07\,$\pm$\,0.04\\
 \hline
 W2  &  1999.25 &  0.12\,$\pm$\,0.02 &  0.19\,$\pm$\,0.12 &  160.03\,$\pm$\,55.28  &  0.12\,$\pm$\,0.06\\
 W2  &  1999.75 &  0.12\,$\pm$\,0.02 &  0.23\,$\pm$\,0.07 &  171.15\,$\pm$\,19.63  &  0.07\,$\pm$\,0.03\\
 W2  &  2001.75 &  0.49\,$\pm$\,0.10 &  0.43\,$\pm$\,0.13 &  170.32\,$\pm$\,20.47  &  0.13\,$\pm$\,0.07\\
 W2  &  2002.75 &  0.09\,$\pm$\,0.02 &  0.47\,$\pm$\,0.04 &  163.98\,$\pm$\,4.68   &  0.04\,$\pm$\,0.02\\
 \hline
 W3  &  2001.75 &  0.25\,$\pm$\,0.05 &  0.15\,$\pm$\,0.04 &  176.11\,$\pm$\,16.78  &  0.04\,$\pm$\,0.02\\
 W3  &  2002.25 &  0.24\,$\pm$\,0.05 &  0.31\,$\pm$\,0.12 &  178.72\,$\pm$\,26.71  &  0.12\,$\pm$\,0.06\\
 W3  &  2002.75 &  0.25\,$\pm$\,0.05 &  0.39\,$\pm$\,0.06 &  175.11\,$\pm$\,9.05   &  0.06\,$\pm$\,0.03\\
 \hline
 W4  &  2002.75 &  0.07\,$\pm$\,0.01 &  0.22\,$\pm$\,0.05 &  164.33\,$\pm$\,15.35  &  0.05\,$\pm$\,0.03\\
 \hline
 W5  &  2002.75 &  0.26\,$\pm$\,0.05 &  0.04\,$\pm$\,0.03 &  -160.75\,$\pm$\,62.06 &  0.03\,$\pm$\,0.01\\
 \hline
 W6  &  2004.75 &  0.56\,$\pm$\,0.11 &  0.07\,$\pm$\,0.05 &  116.79\,$\pm$\,65.90  &  0.05\,$\pm$\,0.03\\
 W6  &  2005.25 &  0.23\,$\pm$\,0.05 &  0.18\,$\pm$\,0.17 &  155.33\,$\pm$\,118.69 &  0.17\,$\pm$\,0.09\\
 W6  &  2007.75 &  1.46\,$\pm$\,0.29 &  0.20\,$\pm$\,0.15 &  172.03\,$\pm$\,77.76  &  0.15\,$\pm$\,0.07\\
 W6  &  2008.33 &  0.71\,$\pm$\,0.14 &  0.28\,$\pm$\,0.16 &  158.93\,$\pm$\,49.09  &  0.16\,$\pm$\,0.08\\
 W6  &  2008.75 &  0.72\,$\pm$\,0.14 &  0.26\,$\pm$\,0.07 &  156.89\,$\pm$\,17.74  &  0.07\,$\pm$\,0.04\\
 \hline
 W7  &  2007.75 &  0.70\,$\pm$\,0.14 &  0.07\,$\pm$\,0.02 &  169.50\,$\pm$\,17.13  &  0.02\,$\pm$\,0.01\\
 W7  &  2008.33 &  0.36\,$\pm$\,0.07 &  0.16\,$\pm$\,0.06 &  144.20\,$\pm$\,26.67  &  0.06\,$\pm$\,0.03\\
 W7  &  2008.75 &  0.54\,$\pm$\,0.11 &  0.17\,$\pm$\,0.04 &  119.22\,$\pm$\,15.17  &  0.04\,$\pm$\,0.02\\
 \hline
 W8  &  2008.33 &  0.37\,$\pm$\,0.07 &  0.13\,$\pm$\,0.04 &  179.03\,$\pm$\,20.07  &  0.04\,$\pm$\,0.02\\
 W8  &  2008.75 &  1.09\,$\pm$\,0.22 &  0.14\,$\pm$\,0.03 &  154.30\,$\pm$\,11.76  &  0.03\,$\pm$\,0.01\\
 W8  &  2010.75 &  0.51\,$\pm$\,0.10 &  0.20\,$\pm$\,0.10 &  -175.97\,$\pm$\,39.25 &  0.10\,$\pm$\,0.05\\
 \hline
 W9  &  2011.33 &  0.02\,$\pm$\,0.00 &  0.08\,$\pm$\,0.02 &  157.51\,$\pm$\,16.56  &  0.02\,$\pm$\,0.01\\
 W9  &  2013.67 &  0.05\,$\pm$\,0.01 &  0.15\,$\pm$\,0.01 &  -179.69\,$\pm$\,6.64  &  0.05\,$\pm$\,0.01\\
 W9  &  2014.33 &  0.36\,$\pm$\,0.07 &  0.20\,$\pm$\,0.01 &  168.14\,$\pm$\,4.26   &  0.05\,$\pm$\,0.01\\
 W9  &  2014.67 &  0.06\,$\pm$\,0.01 &  0.27\,$\pm$\,0.02 &  176.51\,$\pm$\,3.86   &  0.06\,$\pm$\,0.01\\
 \hline
 W10 &  2011.33 &  0.10\,$\pm$\,0.02 &  0.04\,$\pm$\,0.02 &  95.73\,$\pm$\,71.27   &  0.05\,$\pm$\,0.01\\
 W10 &  2012.33 &  0.13\,$\pm$\,0.03 &  0.10\,$\pm$\,0.02 &  95.46\,$\pm$\,20.38   &  0.06\,$\pm$\,0.01\\
 W10 &  2013.67 &  0.27\,$\pm$\,0.05 &  0.04\,$\pm$\,0.01 &  79.48\,$\pm$\,39.22   &  0.04\,$\pm$\,0.01\\
 W10 &  2014.33 &  0.61\,$\pm$\,0.12 &  0.05\,$\pm$\,0.01 &  74.04\,$\pm$\,23.92   &  0.02\,$\pm$\,0.01\\
 W10 &  2014.67 &  1.04\,$\pm$\,0.21 &  0.09\,$\pm$\,0.02 &  87.86\,$\pm$\,17.78   &  0.06\,$\pm$\,0.01\\
 W10 &  2015.33 &  0.23\,$\pm$\,0.05 &  0.12\,$\pm$\,0.02 &  96.81\,$\pm$\,11.30   &  0.05\,$\pm$\,0.01\\
 \hline
 W11 &  2014.33 &  1.15\,$\pm$\,0.23 &  0.08\,$\pm$\,0.01 &  165.62\,$\pm$\,15.53  &  0.06\,$\pm$\,0.01\\
 W11 &  2014.67 &  0.44\,$\pm$\,0.09 &  0.15\,$\pm$\,0.02 &  141.40\,$\pm$\,11.26  &  0.11\,$\pm$\,0.01\\
 W11 &  2015.33 &  0.47\,$\pm$\,0.09 &  0.15\,$\pm$\,0.02 &  157.32\,$\pm$\,14.09  &  0.13\,$\pm$\,0.01\\
 W11 &  2016.67 &  1.21\,$\pm$\,0.24 &  0.16\,$\pm$\,0.02 &  164.22\,$\pm$\,11.06  &  0.14\,$\pm$\,0.01\\
 W11 &  2017.67 &  0.48\,$\pm$\,0.10 &  0.21\,$\pm$\,0.02 &  175.87\,$\pm$\,5.37   &  0.09\,$\pm$\,0.01\\
 \hline
 W12 &  2016.67 &  0.02\,$\pm$\,0.00 &  0.08\,$\pm$\,0.02 &  133.28\,$\pm$\,18.83  &  0.05\,$\pm$\,0.01\\
 W12 &  2017.25 &  0.07\,$\pm$\,0.01 &  0.17\,$\pm$\,0.01 &  151.50\,$\pm$\,5.73   &  0.07\,$\pm$\,0.01\\
 W12 &  2017.67 &  0.18\,$\pm$\,0.04 &  0.23\,$\pm$\,0.02 &  142.48\,$\pm$\,4.36   &  0.07\,$\pm$\,0.01\\
 \hline
 W13 &  2017.67 &  1.26\,$\pm$\,0.25 &  0.12\,$\pm$\,0.02 &  92.64\,$\pm$\,11.13   &  0.07\,$\pm$\,0.01\\
 W13 &  2019.25 &  0.66\,$\pm$\,0.13 &  0.14\,$\pm$\,0.02 &  154.19\,$\pm$\,8.46   &  0.07\,$\pm$\,0.01\\
 \hline
 W14 &  2019.25 &  0.43\,$\pm$\,0.09 &  0.05\,$\pm$\,0.02 &  30.50\,$\pm$\,28.16   &  0.03\,$\pm$\,0.01\\
 W14 &  2020.25 &  0.62\,$\pm$\,0.12 &  0.07\,$\pm$\,0.01 &  84.37\,$\pm$\,13.39   &  0.03\,$\pm$\,0.01\\
 W14 &  2020.75 &  0.48\,$\pm$\,0.10 &  0.16\,$\pm$\,0.01 &  74.03\,$\pm$\,7.11    &  0.07\,$\pm$\,0.01\\
 \hline
 W15 &  2020.25 &  0.06\,$\pm$\,0.01 &  0.21\,$\pm$\,0.01 &  170.47\,$\pm$\,4.12   &  0.04\,$\pm$\,0.01\\
 W15 &  2020.75 &  0.10\,$\pm$\,0.02 &  0.23\,$\pm$\,0.01 &  164.54\,$\pm$\,3.79   &  0.04\,$\pm$\,0.01\\
\hline
\end{tabular}
\end{table*}

Table~\ref{table:Vel4386} contains a list of the IDs and velocities of the cross-identified features.
For these cross-identified features, we estimate the error for the ejection year to be $\pm6$ months, which is set by the average cadence of available 86\,GHz epochs.
Tables \ref{table:Vel86} and \ref{table:Vel43} display the feature IDs and corresponding interpolated velocity, at 86 and 43\,GHz respectively.
Components ``W4'', ``W5'', and ``W13'' were only identified in one epoch and we therefore could not derive a velocity estimate.
For the individual features, we impose a $5\sigma$ uncertainty (with $1\sigma=0.5$\,years, from the cross-identified features), and thus conservatively estimate the error for the ejection year to be $\pm2.5$ years.
A description of the fitting method and error estimation is offered in Appendix \ref{App:Comp_vel_mod}.

\begin{table}[h]
\caption{Combined feature velocity and ejection date of cross-identified features.}        
\label{table:Vel4386}    
\centering  
\begin{tabular}{ccccc}
ID & ID$_{43}$ & ID$_{86}$ & Ejection date & Velocity (c)\\
\hline\hline
 F1 &  Q1 &  W6  &  2001.96\,$\pm$\,0.5 &  0.047\,$\pm$\,0.004\\
 F2 &  Q2 &  W9  &  2003.50\,$\pm$\,0.5  &  0.034\,$\pm$\,0.004\\
 F3 &  Q3 &  W8  &  2011.79\,$\pm$\,0.5 &  0.066\,$\pm$\,0.005\\
 F4 &  Q4 &  W12 &  2015.86\,$\pm$\,0.5 &  0.117\,$\pm$\,0.007\\
 F5 &  Q6 &  W15 &  2017.88\,$\pm$\,0.5 &  0.115\,$\pm$\,0.010 \\
\hline
\end{tabular}
\end{table}

\begin{table}[h]
\caption{Component velocity and ejection date at 86\,GHz.}       
\label{table:Vel86}     
\centering  
\begin{tabular}{ccc}
ID & Ejection date (year) & Velocity (c)\\
\hline\hline
 W1  &  1981.63\,$\pm$\,2.5 &  0.029\,$\pm$\,0.011\\
 W2  &  1996.86\,$\pm$\,2.5 &  0.100\,$\pm$\,0.008  \\
 W3  &  2001.15\,$\pm$\,2.5 &  0.312\,$\pm$\,0.039\\
 W4  &  --                  &  --                 \\
 W5  &  --                  &  --                 \\
 W6  &  2002.72\,$\pm$\,2.5 &  0.054\,$\pm$\,0.008\\
 W7  &  2006.89\,$\pm$\,2.5 &  0.110\,$\pm$\,0.024 \\
 W8  &  1993.38\,$\pm$\,2.5 &  0.012\,$\pm$\,0.010 \\
 W9  &  2010.15\,$\pm$\,2.5 &  0.061\,$\pm$\,0.020 \\
 W10 &  2007.79\,$\pm$\,2.5 &  0.013\,$\pm$\,0.013\\
 W11 &  2011.08\,$\pm$\,2.5 &  0.038\,$\pm$\,0.010 \\
 W12 &  2016.13\,$\pm$\,2.5 &  0.186\,$\pm$\,0.002\\
 W13 &  --                  &  --                 \\
 W14 &  2018.64\,$\pm$\,2.5 &  0.078\,$\pm$\,0.042\\
 W15 &  --                  &  --                 \\
\hline
\end{tabular}
\end{table}

\begin{table}[h]
\caption{Component velocity and ejection date at 43\,GHz.}  
\label{table:Vel43}   
\centering  
\begin{tabular}{ccc}
ID & Ejection date (year) & Velocity (c)\\
\hline\hline
 Q1 &  2001.87\,$\pm$\,2.5 &  0.05\,$\pm$\,0.02\\
 Q2 &  2011.30\,$\pm$\,2.5 &  0.06\,$\pm$\,0.01\\
 Q3 &  2008.92\,$\pm$\,2.5 &  0.07\,$\pm$\,0.05\\
 Q4 &  2015.84\,$\pm$\,2.5 &  0.12\,$\pm$\,0.01\\
 Q5 &  --                  &  --               \\
 Q6 &  2017.88\,$\pm$\,2.5 &  0.11\,$\pm$\,0.01\\
\hline
\end{tabular}

\end{table}

\begin{figure*}
  \centering
  \includegraphics[scale=0.35]{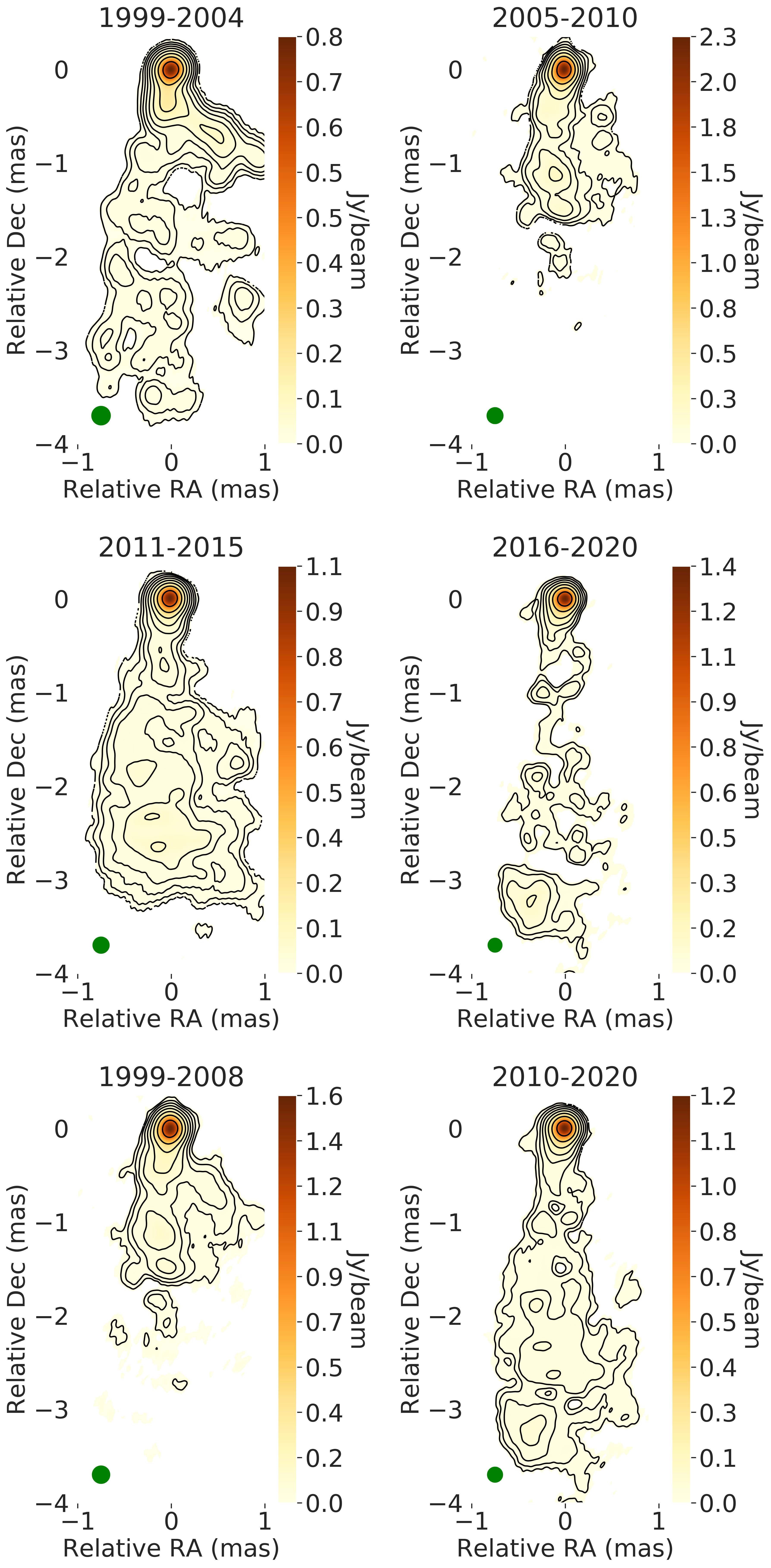}
      \caption{Full showcase of stacked images of  \C\ at 86\,GHz, with different time bins. 
      The title of each panel indicates the time bin used.
      The same procedure was followed to produce the panels, as described in the caption of Fig.~\ref{fig:Individual_Stacked}.
      All image parameters are summarised in Table~\ref{table:CircBeamBin}}
         \label{fig:Full86}
  \end{figure*}

Table~\ref{table:CircBeam} displays the circular convolving beam used to produce the stacked images of Fig.~\ref{fig:Individual_Stacked}, per frequency.
The last two columns correspond to the intensity peak of the stacked image, and the rms error.
Similarly, Table~\ref{table:CircBeamBin} displays the parameters associated with Fig.~\ref{fig:Full86}.

\begin{table}[h]
\caption{Summary of image parameters for the stacked images.}   
\label{table:CircBeam}   
\centering   
\begin{tabular}{ccccc}
Frequency (GHz) & Beam radius (mas) & Pixel scale (mas/pixel) & S$_\text{max}$ (Jy/beam) & rms (mJy/beam)\\
\hline\hline
 15.4 &  0.56 &  0.03 &  3.18$\,\pm\,$0.64 &  0.45\\
 43.1 &  0.24 &  0.02 &  3.02$\,\pm\,$0.60 &  0.24\\
 86.3 &  0.08 &  0.01 &  0.75$\,\pm\,$0.15 &  0.77\\
\hline
\end{tabular}
\end{table}

\begin{table*}[h]
\caption{Summary of image parameters of the spectral index maps.}
\label{table:SPI}      
\centering   
\begin{tabular}{ccc}
Epoch & Beam radius (mas) & Pixel scale (mas/pixel)\\
\hline\hline
 Stacked &  0.24 &  0.02\\
 2011.33 &  0.22 &  0.02\\
 2014.33 &  0.25 &  0.02\\
 2014.67 &  0.21 &  0.02\\
 2017.25 &  0.22 &  0.02\\
 2017.67 &  0.22 &  0.02\\
 2020.75 &  0.28 &  0.02\\
\hline
\end{tabular}
\end{table*}

\begin{table}[h]
\caption{Summary of image parameters of the time binned stacked images.}     
\label{table:CircBeamBin} 
\centering  
\begin{tabular}{ccccc}
Time bin (years) & Beam radius (mas) & Pixel scale (mas/pixel) & S$_\text{max}$ (Jy/beam) & rms (mJy/beam)\\
\hline\hline
 1999-2004 &  0.20 &  0.01 &  0.81$\,\pm\,$0.16 &  0.25\\
 2005-2010 &  0.17 &  0.01 &  2.31$\,\pm\,$0.46 &  1.82\\
 2011-2015 &  0.17 &  0.01 &  1.06$\,\pm\,$0.21 &  0.33\\
 2016-2020 &  0.15 &  0.01 &  1.38$\,\pm\,$0.28 &  1.61\\
 1999-2008 &  0.19 &  0.01 &  1.60$\,\pm\,$0.32 &  0.92\\
 2010-2020 &  0.16 &  0.01 &  1.23$\,\pm\,$0.25 &  0.83\\
\hline
\end{tabular}
\end{table}

Finally Table~\ref{table:Gamma} displays the power law indices per frequency.
The fitting procedure and error estimation is explained in Appendix \ref{App:ImStack}.

\begin{table}[h]
\caption{Power law fit parameters for the stacked images, per frequency.}             
\label{table:Gamma}     
\centering  
\begin{tabular}{cc}
Frequency (GHz) & Power law index\\
\hline\hline
 15.4 &  0.36$\,\pm\,0.38$\\
 43.1 &  0.27$\,\pm\,0.13$\\
 86.3 &  0.90$\,\pm\,0.06$\\
\hline
\end{tabular}
\end{table}

\end{appendix}
\end{document}